\documentclass[12pt]{iopart}

\usepackage[utf8]{inputenc}

\usepackage{iopams}

\expandafter\let\csname equation*\endcsname\relax
\expandafter\let\csname endequation*\endcsname\relax
\usepackage[fleqn]{amsmath}

\usepackage{graphicx}
\usepackage{subfig}

\usepackage{array}
\usepackage{multirow}

\newcommand{\bra}[1]{\langle#1|}
\newcommand{\ket}[1]{|#1\rangle}

\begin{document}

\title{Semiconductor devices for entangled photon pair generation: a review}

\author{Adeline Orieux$^{1,2}$, Marijn A. M. Versteegh$^{3}$, Klaus D. Jöns$^{3}$, and Sara Ducci$^{4}$}

\address{$^1$ Sorbonne Universit\'es, UPMC Univ Paris 06, CNRS, Laboratoire d'Informatique de Paris 6 (LIP6), 4 Place Jussieu, 75005 Paris, France.}
\address{$^2$ IRIF UMR 8243, Universit\'e Paris Diderot, Sorbonne Paris Cit\'e, CNRS, 75013 Paris, France.}
\address{$^3$ Department of Applied Physics, Quantum Nano Photonics Group, Royal Institute of Technology (KTH), Stockholm 106 91, Sweden.}
\address{$^4$ Laboratoire Mat\'eriaux et Ph\'enom\`enes Quantiques, Universit\'e Paris Diderot, Sorbonne Paris Cit\'e, CNRS-UMR 7162, 75205 Paris Cedex 13, France.}
\ead{sara.ducci@univ-paris-diderot.fr}
\ead{klausj@kth.se}

\vspace{10pt}
\begin{indented}
\item February 2017
\end{indented}

\begin{abstract}
Entanglement is one of the most fascinating properties of quantum mechanical systems; when two particles are entangled the measurement of the properties of one of the two allows to instantaneously know the properties of the other, whatever the distance separating them. In parallel with fundamental research on the foundations of quantum mechanics performed on complex experimental set-ups, we assist today to a bourgeoning of quantum information technologies bound to exploit entanglement for a large variety of applications such as secure communications, metrology and computation. Among the different physical systems under investigation, those involving photonic components are likely to play a central role and in this context semiconductor materials exhibit a huge potential in terms of integration of several quantum components in miniature chips. In this article we review the recent progress in the development of semiconductor devices emitting entangled photons. We will present the physical processes allowing to generate entanglement and the tools to characterize it; we will give an overview of major recent results of the last years and highlight perspectives for future developments.
\end{abstract}


\tableofcontents


\section{Introduction}

Entanglement is one of the weirdest and most fascinating properties of quantum systems. This concept, whose name was invented by Schrödinger~\cite{Schrodinger1935} was at the center of a famous paper published by Einstein, Podolsky and Rosen in 1935~\cite{EPR1935}. In this work, the authors analysed the predictions of correlation measurements for a two-particle state, where neither particle can be considered in a state independent from the other, but form instead a single entangled system.
They made two assumptions. The first assumption (later called the assumption of `realism') is that `if, without in any way disturbing a system, we can predict with certainty (i.e., with probability equal to unity) the value of a physical quantity, then there exists an element of physical reality corresponding to this physical quantity.' The second assumption (called `locality')  is that a measurement performed on one particle cannot influence the properties of the other one when the particles no longer interact (for example when they have been brought to a large distance from each other). Based on these two assumptions, Einstein, Podolsky and Rosen argued that the description of reality as given by the laws of quantum mechanics is not complete. This argument arose fierce debates among the founders of quantum mechanics, and became experimentally testable with Bell's discovery of the so-called Bell inequalities in 1964~\cite{Bell1964} and their extension to experimental conditions by Clauser \textit{et al.}~\cite{Clauser1969,Clauser1974}. These Bell inequalities show that, for certain combinations of measurement settings, quantum mechanics predicts correlations between the outcomes of measurements performed on the two particles that are incompatible with the joint assumption of realism and locality. Starting from the '70s~\cite{Freedman1972} and early '80s~\cite{Aspect1981} several generations of more and more refined experiments have been implemented to test Bell's inequalities and falsify local realism. At the same time, the existence of entangled particles was demonstrated over larger and larger distances, with an actual record of more than $300\,km$~\cite{Korzh2015}.

\noindent Yet, despite their ingenuity, the performed Bell tests were not perfect in the sense that additional assumptions were required (the so-called `loopholes'), which could allow one to maintain local realism and explain the observed correlations by some other effect. Recently, four experiments have closed all significant loopholes simultaneously~\cite{Hensen2015,Giustina2015,Shalm2015,Rosenfeldarxiv}.
Thus, these experiments convincingly demonstrated that quantum entanglement exists, and that nature cannot be described by any local realistic theory, that is a theory where physical properties exist independently of measurement and where there is no physical influence faster than light. 
Nevertheless, there is still an interest in performing more entanglement tests for several reasons. A first motivation is the exploration of the boundary between the quantum and the classical world, which is a subject of intense research both for theoreticians and experimentalists~\cite{Brunner2014}. A second motivation is provided by experiments which extend the traditional set-up for Bell-type tests to relativistic configurations and investigate the so-called relativistic non-locality~\cite{Suarez1997}. Indeed, several groups all around the world are involved in a sort of ‘quantum space race’ consisting in sending satellites equipped with quantum technologies into space to test fundamental physics in new regimes~\cite{Jennewein2013,ChineseSatellite}. 

Apart from these fundamental motivations, in these last 30 years, we have assisted to the booming of a new field, namely quantum information science, whose objective is to enable new forms of communication, computation and measurement based on the utilization of quantum mechanical systems~\cite{Jaeger2007}, with entanglement playing a central role. Quantum information is both a fundamental science, gathering together specialists of different disciplines (physics, mathematics, informatics, material science,..), and a progenitor of novel technologies, as witnessed by the number of companies working in this field that have emerged over the last years. In particular, several commercial quantum key distribution systems are already available, offering enhanced security by using cryptographic keys encoded in quantum systems~\cite{Gisin2002}. A long-term anticipated future technology is the quantum computer~\cite{Ladd2010,Castelvecchi2017}, which should work exponentially faster than its classical counterpart for particular tasks and could enable the simulation of complex quantum systems. Quantum metrology, which aims at achieving the highest precision allowed in nature by exploiting quantum effects in the measurement process~\cite{Giovanetti2004,Humphreys2013}, has also attracted a lot of research efforts.
Maybe the most appealing application for photons is long-distance quantum communication~\cite{Zoller2001}. Indeed, photons naturally behave as flying qubits, able to travel at the speed of light over long distances, and are almost immune to decoherence. Thus, they are a key ingredient of the future so-called `quantum internet'~\cite{Kimble2008} which is envisoned as a network of quantum links, over which photons will transport quantum information, and quantum nodes, consisting of solid-state or atomic systems that will process or relay this information. The benefits of such a quantum network would be manyfold, such as e.g. the unconditional security of information exchanges enabled by quantum cryptography~\cite{Gisin2002,Scarani2009,Lo2014}, the possibility of secure delegated quantum cloud computing~\cite{Childs2005,Barz2012}, and ways of achieving some communication tasks that are not permitted or are less efficient by classical means~\cite{Arrazola2014}.\\
For this quantum network to work and be deployed on a large scale, practical, reliable and cost-effective quantum components are needed, in particular sources of entangled photons. This is one of the reasons why integrated quantum photonics has been attracting a growing interest in these last years~\cite{OBrien2009,Orieux2016}. In particular, semiconductor materials, which are already at the basis of current classical communication and computation technologies, are an ideal platform for the miniaturization and integration of several quantum components, opening the way to the generation, manipulation and detection of quantum states of light on a same chip.

\vspace{0.5cm}
In this article, we review semiconductor devices for the generation of entangled photons, addressing both fundamental and applied aspects. Although there has been considerable progress in experiments based on continuous quantum variables with semiconductor devices as well~\cite{Masada2015}, here we will focus only on entanglement between discrete two-level quantum systems (called quantum bits or qubits). 
The article is structured along the following lines: in Section~2, we introduce the concept of qubits, entangled states and the physical processes used to generate two-photon entanglement with semiconductor devices. In Section~3, we present the main methods used to measure entanglement. Section~4 gives a review of the current state of the art of semiconductor devices generating photonic entanglement. Finally, Section~5 is an opening on recent applications and prospects.


\section{Photonic qubits and entanglement generation}
\subsection{Photonic Qubits}

The basic entity in classical information theory is the bit, which can take either of two values: 0 or 1. Its quantum analog, the quantum bit or `qubit', is a two-dimensional quantum system whose basic states $\ket{0}$ and $\ket{1}$ form an orthogonal basis of the qubit space, called the computational basis. Unlike the classical bit, the qubit can be in a coherent superposition of $\ket{0}$ and $\ket{1}$, its general (pure) state being:
\begin{equation}
\fl
\ket{\psi_{qubit}}=\alpha\ket{0}+\beta \textrm{e}^{i\phi}\ket{1},
\label{eqQubit}
\end{equation}
with $\alpha^2+\beta^2=1$.\\
This means that the outcome of the measurement of a qubit is not always deterministic: for the state defined above, a measurement in the computational basis will give the result 0 or 1 with a probability $\alpha^2$ or $\beta^2$ respectively. Note that this could still be achieved with a classical bit in a statistical mixture between $\ket{0}$ and $\ket{1}$, however the  unique feature of a qubit is that the basic states are superposed coherently, a difference that can be evidentiated by a measurement in a different basis. The interaction of a qubit with its environment in a thermodynamically irreversible way can cause a transition from a fully coherent superposition (pure state) to an incoherent one (mixed state): this process is called decoherence.

\begin{figure}[htb]
\centerline{\includegraphics[width=0.65\textwidth]{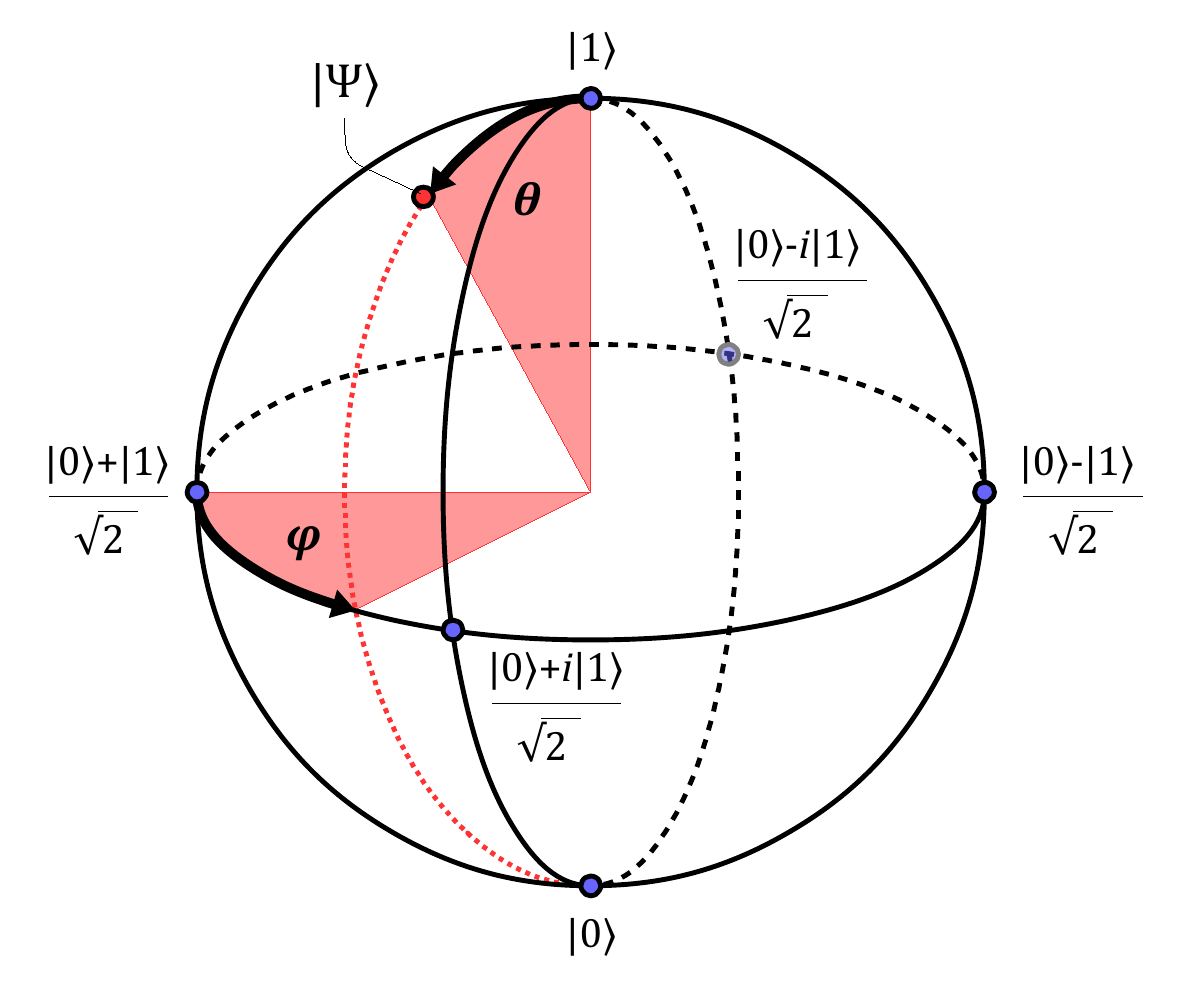}}
\caption{The Bloch sphere. A pure state $\ket{\Psi}=\alpha\ket{0}+\beta \textrm{e}^{i\phi}\ket{1}$ corresponds to a point on the sphere with spherical coordinates ($\varphi = 2\phi$ ; $\theta = 2\arctan{(\alpha/\beta)}$). Mixed states are inside the sphere.}
\label{SD1}
\end{figure}

Qubits can be represented graphically on the qubit sphere, also called the Bloch sphere (see Figure~\ref{SD1}). The basic states $\ket{0}$ and $\ket{1}$ are located in the poles of the sphere. Any two diametrically opposed points on the sphere correspond to two orthogonal states that form an orthogonal basis. All pure states (i.e. written in the form of Equation~\ref{eqQubit}) are situated on the surface of the sphere. The azimutal angle $\varphi$ is related to the phase $\phi$ ($\varphi = 2\phi$) while the polar angle $\theta$ is related to the coefficients $\alpha$ and $\beta$ ($\theta = 2\arctan{(\alpha/\beta)}$). Points on the equator correspond to pure states with equal coefficients $\alpha$ and $\beta$. Mixed states are found inside the sphere, the center of the sphere corresponding to a completely mixed state.

Photonic qubits can be obtained by exploiting different properties of single photons; the available degrees of freedom being the photons’ polarization, spatial mode, temporal mode, orbital angular momentum mode and frequency. The spectral range chosen for the photons will depend on their intended use. In particular, if they have to be transmitted through optical fibres, as is often requested for long-distance quantum communication applications, they should have a wavelength in the second or third telecommunication window (around $1319$ or $1555\,nm$, respectively). For each spectral region of the electromagnetic field, different kinds of single-photon detectors have been developed. Note that single-photon detection is a very active field of research in itself, having pushed forward several technologies and gathering the interest of several communities (astronomers, biologists, medical researchers, quantum physicists,…).\\
In the following, we describe different kinds of photonic qubits. Following the vocabulary of quantum communications, we call Alice the party who prepares the qubit (also known as the sender) and Bob the one who measures it (the receiver).

The most well-known realization of a qubit is the \textbf{polarization qubit}, which consists of orthogonal states of polarization. In the qubit sphere, we can identify left and right circularly polarized photons with the computational basis states $\ket{0}$ and $\ket{1}$; they correspond to the poles of the sphere. Linearly polarized states can be found on the equator, and elliptically polarized light everywhere else on the sphere. Polarization qubits can be very easily created and measured using polarizers and waveplates oriented at arbitrary angles.

\begin{figure}[htb]
\centerline{\includegraphics[width=0.8\textwidth]{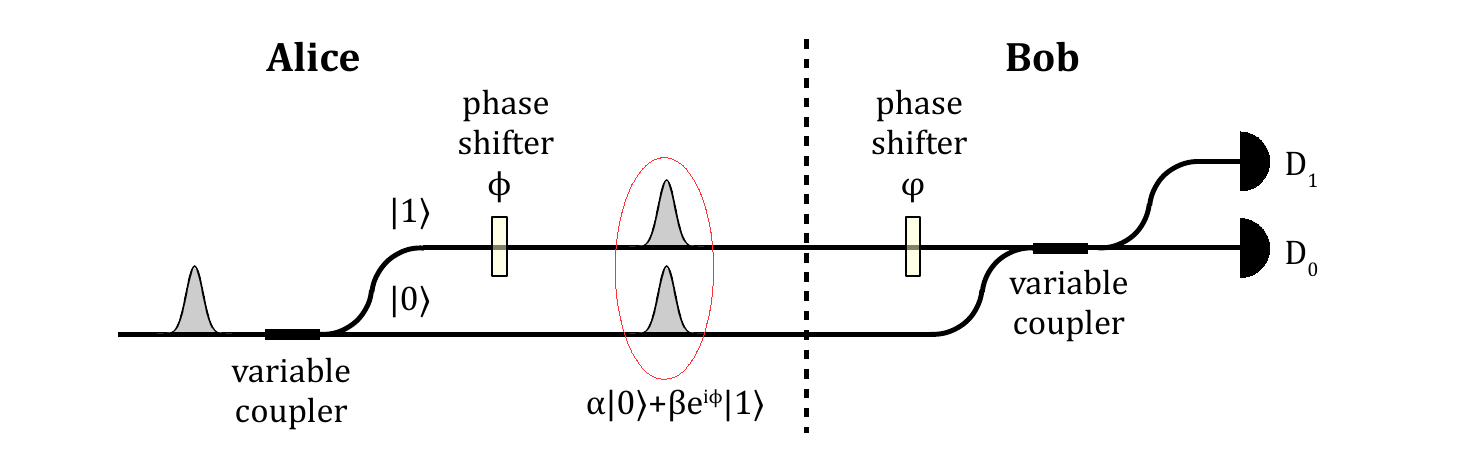}}
\caption{Creation (on Alice's side) and measurement (on Bob's side) of a spatial mode qubit. D$_0$ and D$_1$ are single-photon detectors. The variable coupler is used to adjust the values of $\alpha$ and $\beta$, Alice's phase shifter sets the phase $\phi$ and Bob's phase shifter sets the measurement basis.}
\label{SD2}
\end{figure}

\noindent Another possibility is the \textbf{spatial mode qubit}, shown in Figure~\ref{SD2}, also known as the `dual-rail' qubit. Here, the states $\ket{0}$ and $\ket{1}$ correspond to two possible propagation modes. Alice can create any desired superposition by using a variable coupler and a phase shifter. A similar set-up can be used by Bob to analyse the qubit. Note that in these last years, the emergence of integrated photonic technology in the realm of quantum applications has given rise to a new generation of integrated waveguide structures consisting of complex quantum circuits with intrinsic phase stability, perfectly adapted to this type of qubit.

\begin{figure}[htb]
\centerline{\includegraphics[width=0.8\textwidth]{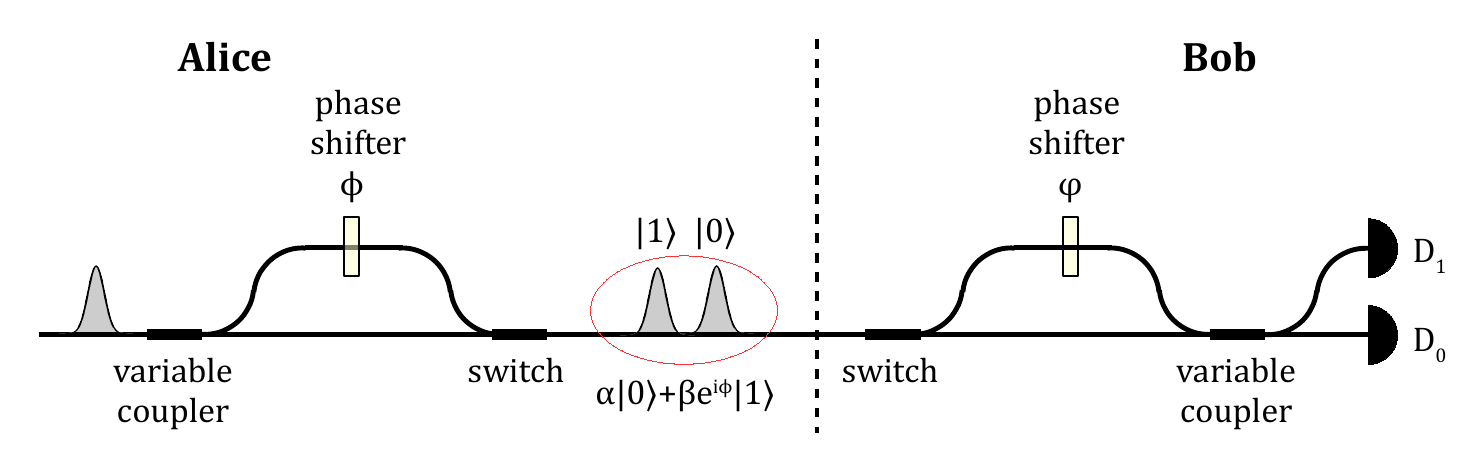}}
\caption{Creation (on Alice's side) and measurement (on Bob's side) of a time-bin qubit. On Alice's side, the variable coupler is used to adjust the values of $\alpha$ and $\beta$ and the phase shifter sets the phase $\phi$. On Bob's side, the phase shifter and the variable coupler are used to set the measurement basis.}
\label{SD3}
\end{figure}

\noindent Figure~\ref{SD3} shows the scheme for the realization of a so-called \textbf{time-bin qubit}. After the separation of the photon into two spatial modes with a variable coupler, Alice uses a switch to transfer the amplitudes of both spatial modes -- arriving on the switch with a time-difference much larger than the photon's coherence time -- back into the same spatial mode. In this way Alice creates a superposition of amplitudes describing a photon in two different time-bins. To undo this transformation and measure the qubit, Bob uses a symmetrical set-up.

\noindent In the last decade, another property of light has attracted the attention of the community: orbital angular momentum (OAM). This property is related to the photon's transverse-mode spatial structure; the eigenvalues of the orbital angular momentum operator can be any positive or negative integer value, that physically refers to the number of twistings of the phase along the propagation direction in clockwise (positive) or counter-clockwise (negative) orientation. The recent progress in the generation and manipulation of OAM has led to the demonstration of \textbf{OAM qubits} and entanglement of the orbital angular momentum states of photons~\cite{Mair2001}.

\noindent \textbf{Frequency qubits} can also be created by using a superposition of basic states at frequencies $\omega_1$ and $\omega_2$, as it is done with atoms; this approach has not been well developed yet, mainly because of the difficulty in chosing arbitrary measurement basis for frequency, nevertheless we can cite some interesting experimental works recently done with these qubits~\cite{Olislager2010,Olislager2014}.

The examples cited above are restricted to two-dimensional Hilbert spaces. This restriction, however, is strict only for the polarization degree of freedom: spatial and temporal modes, frequencies and orbital angular momenta, on the other hand, can be described by Hilbert spaces of much higher dimensions, thus giving the possibility of encoding `qu$d$its' (i.e. quantum states of dimension $d>2$). Indeed the realization and manipulation of superpositions in higher dimensions is an active field of research in quantum information science, offering further capabilities for quantum information processing, in particular quantum computation with reduced requirements in the number of interacting quantum particles~\cite{Sanders2002}.

\subsection{Two-qubit entanglement}

In this section, we show some of the most common ways of generating entangled states of two photons in the different degrees of freedom. For the interested reader, we suggest two review papers on entanglement: a theoretical one~\cite{Horodecki2009} and an experiment-oriented one~\cite{Edamatsu2007}.
Entanglement can be seen as the generalization of the superposition principle to multi-particle systems.
The state describing the whole of an entangled multi-particle system cannot be factorized, i.e. written as a tensor product of the properties associated with each subsystem. For example, two-qubit entangled pure states can be written as:
\begin{equation}
\fl
\ket{\psi}=\alpha\ket{0}_A\ket{0}_B+\beta e^{i\phi}\ket{1}_A\ket{1}_B,
\end{equation}
or
\begin{equation}
\fl
\ket{\psi}=\alpha\ket{0}_A\ket{1}_B+\beta e^{i\phi}\ket{1}_A\ket{0}_B,
\end{equation}
where the indices $A$ and $B$ label the two photons, and $\alpha^2+\beta^2 = 1$. For $\alpha = \beta$  and $\phi=0,\pi$ we obtain the four well-known Bell states:
\begin{equation}
\fl
\ket{\Phi^\pm}=\ket{0}_A\ket{0}_B\pm\ket{1}_A\ket{1}_B,
\end{equation}
and
\begin{equation}
\fl
\ket{\Psi^\pm}=\ket{0}_A\ket{1}_B\pm\ket{1}_A\ket{0}_B.
\end{equation}

In order to produce entangled photon pairs, there must be two possible and indistinguishable ways of creating such pairs. This can occur either within the source itself or with a post-manipulation through additional optics after the source and post-selection at the detectors. According to the degree of freedom chosen to encode the qubits, different types of entanglement can be generated.\\

\begin{figure}[htb]
\centerline{\includegraphics[width=1\textwidth]{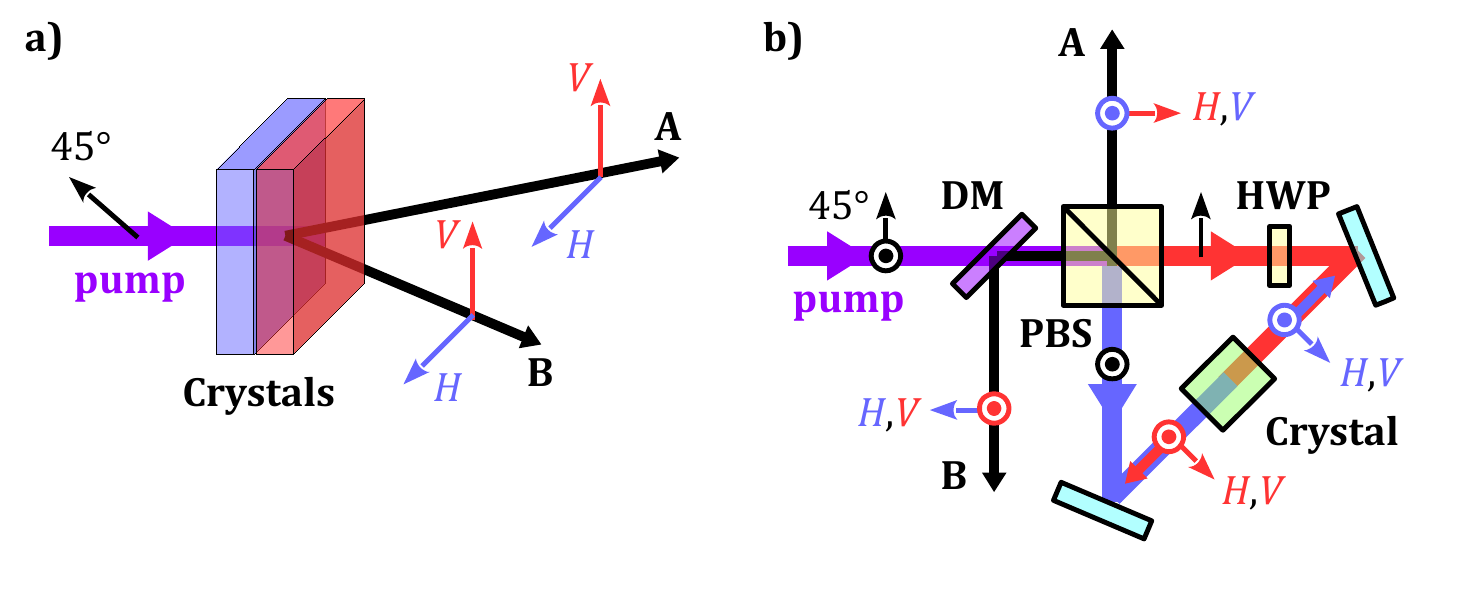}}
\caption{Two examples of schemes to generate polarization-entangled photon states with nonlinear crystals (see the text for details). The thin arrows indicate the polarization directions of the different beams. a) Scheme of Ref.~\cite{Kwiat1999}: the entangled state $\big(\ket{H}_A\ket{H}_B+e^{i\phi}\ket{V}_A\ket{V}_B\big)/\sqrt{2}$ is generated. b) Scheme of Ref.~\cite{Kim2006,Fedrizzi2007,Pryde2016,Giustina2015}: the entangled state $\big(\ket{H}_A\ket{V}_B+e^{i\phi}\ket{V}_A\ket{H}_B\big)/\sqrt{2}$ is generated. DM is a dichroic mirror that transmits the pump beam and reflects the parametric photons. The half-wave plate (HWP) is used to transform a vertical polarization into a horizontal one and vice versa for both the pump beam and the generated photon pairs.}
\label{FigPolEnt}
\end{figure}

Most experiments to date have generated \textbf{polarization entanglement}, because of the easy manipulation of these qubits with polarizers and waveplates and their relatively easy generation with nonlinear crystals~\cite{Shih1988,Ou1988,Kwiat1995,Kwiat1999,Kim2006} (the physical process of nonlinear parametric creation of photon pairs will be explained in Section~2.3.1). Figure~\ref{FigPolEnt} shows two of the currently most popular ways of producing polarization-entangled Bell states in quantum optics laboratories. In Figure~\ref{FigPolEnt}a~\cite{Kwiat1999}, two type-I nonlinear crystals (i.e. generating photons having the same polarization state) with their optical axis orthogonal to each other are pumped by a laser beam (in purple) polarized at 45$^\circ$ with respect to the horizontal axis. The first crystal (in blue) can generate horizontally-polarized photon pairs $\ket{H}_A\ket{H}_B$ while the second crystal (in red) can generate vertically-polarized photon pairs $\ket{V}_A\ket{V}_B$. The probability of emission of a pair being very low, only one crystal generates a pair at a given time but there is no way of knowing which one, provided that the crystals are thin enough so as to ensure that the emitted modes from both crystals are indistinguishable. Thus the photon pair is emitted in a coherent superposition of both states, i.e. in the maximally entangled state $\big(\ket{H}_A\ket{H}_B+e^{i\phi}\ket{V}_A\ket{V}_B\big)/\sqrt{2}$. In Figure~\ref{FigPolEnt}b~\cite{Kim2006,Fedrizzi2007,Pryde2016,Giustina2015}, a type-II nonlinear crystal  (i.e. generating photons having orthogonal polarization states) is inserted in a Sagnac interferometer formed by two mirrors and a polarizing beam splitter (PBS). The pump beam (in purple) is polarized at 45$^\circ$ with respect to the horizontal axis and impinges on the PBS which transmits its horizontal polarization component and reflects the vertical one. In both clockwise (in red) and anticlockwise (in blue) directions, the crystal can emit a $\ket{H}\ket{V}$ photon pair, but only one of these indistinguishable processes occur at a given time. After the PBS, the resulting state of the photon pairs is the maximally entangled state $ \big(\ket{H}_A\ket{V}_B+e^{i\phi}\ket{V}_A\ket{H}_B\big)/\sqrt{2}$. In both schemes, the phase $\phi$ in the generated two-photon state can be modified by adjusting the phase difference between the horizontal and vertical components of the polarization of the pump beam.

\begin{figure}[htb]
\centerline{\includegraphics[width=1\textwidth]{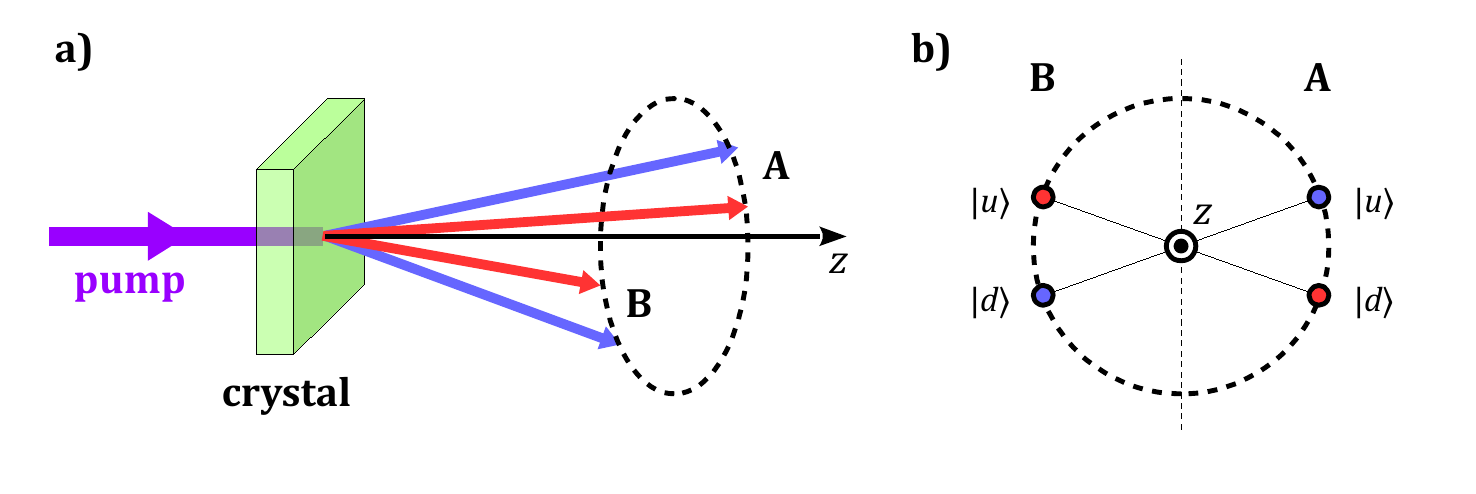}}
\caption{a) Example of a set-up for the generation of momentum-entangled photon pairs~\cite{Barbieri2005} (see the text for details). $z$ is the propagation axis of the pump beam. The entangled state $\big(\ket{u}_A\ket{d}_B+\ket{d}_A\ket{u}_B\big)/\sqrt{2}$ is created. b) Selection of two pairs of spatial modes corresponding to two possible photon pairs.}
\label{FigEntMom}
\end{figure}

\noindent \textbf{Momentum} or \textbf{spatial mode entanglement} is also often used, either with sources based on nonlinear crystals in non-collinear geometries or, more recently, in integrated photonic circuits. In Figure~\ref{FigEntMom}, we show one example of a set-up that can be used to generate momentum-entangled photon pairs~\cite{Barbieri2005}. A laser beam (in purple) pumps a type-I nonlinear crystal. Photon pairs can be emitted with many possible propagation directions all around a cone, the two photons of a pair always pointing in diametrically opposed directions. By selecting two such pairs of directions, an entangled state $\big(\ket{u}_A\ket{d}_B+\ket{d}_A\ket{u}_B\big)/\sqrt{2}$ can be created, where $\ket{u}$ and $\ket{d}$ correspond to the `up' and `down' modes respectively.

\begin{figure}[htb]
\centerline{\includegraphics[width=1\textwidth]{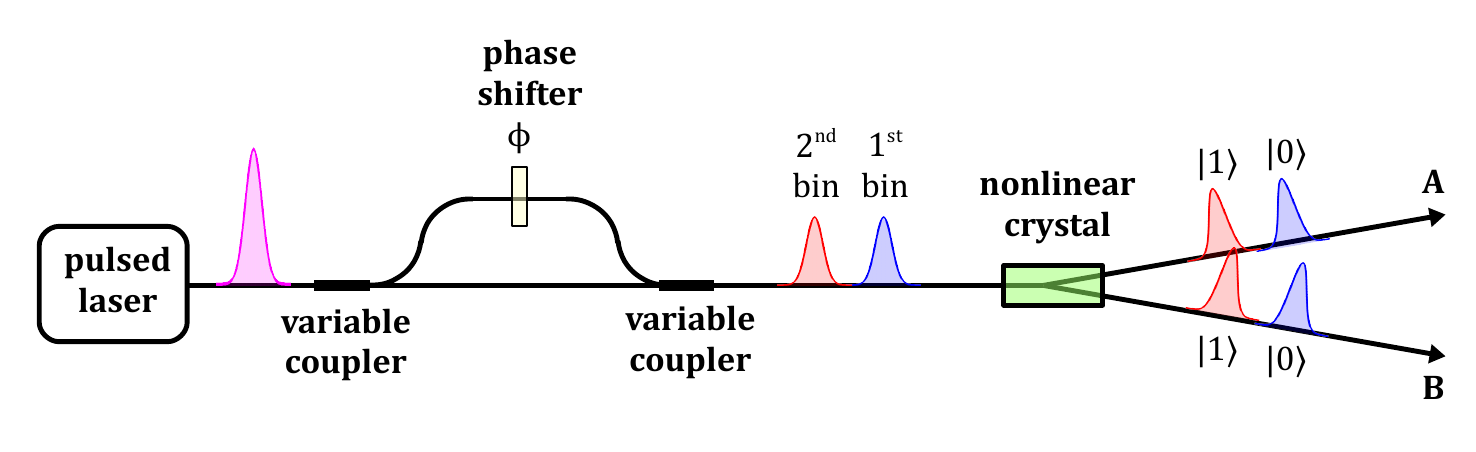}}
\caption{Example of a set-up for the generation of time-bin-entangled photon pairs. An entangled state of the form $\alpha\ket{0}_A\ket{0}_B+\beta e^{i\phi}\ket{1}_A\ket{1}_B$ is generated, where $\alpha$ and $\beta$ are set by the variable couplers and $\phi$ by the phase-shifter.}
\label{FigEntTB}
\end{figure}

\noindent For applications such as long-distance quantum communications in optical fibres, time-bin and energy-time entangled photons have proved to be more robust than the previous two against decoherence effects occuring during the propagation of the photons in the fibre.\\
\textbf{Time-bin entanglement} can be generated using the set-up of Figure~\ref{FigEntTB}~\cite{Zbinden1999}: a classical light-pulse emitted by a pulsed laser is split into two subsequent pulses (or bins) by means of an interferometer with a large path-length difference. This two-pulse laser light is then used to pump a nonlinear crystal in which a photon pair can be created either by the first pulse (in the time-bin $\ket{0}$) or by the second pulse (in the time-bin $\ket{1}$). Depending on the coupling ratios of the couplers of the interferometer and the phase $\phi$, any entangled state of the form $\alpha\ket{0}_A\ket{0}_B+\beta e^{i\phi}\ket{1}_A\ket{1}_B$ can be generated.

\begin{figure}[htb]
\centerline{\includegraphics[width=1\textwidth]{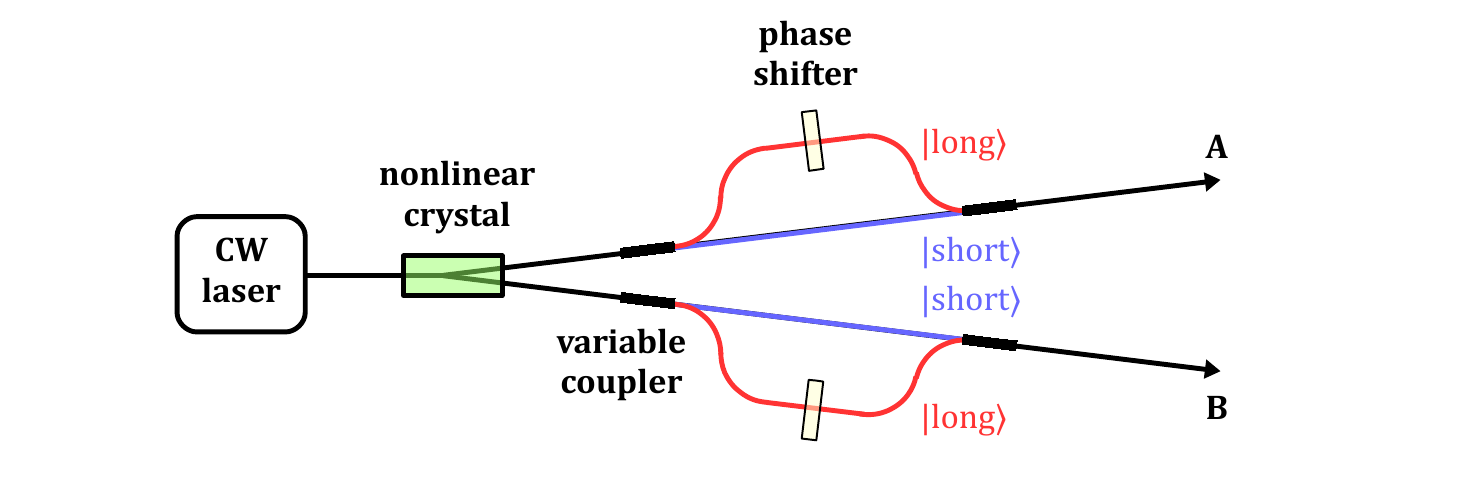}}
\caption{Franson-type set-up for energy-time entanglement (see the text for details.) The photons A and B emitted by a nonlinear crystal are each sent in an interferometer similar to a time-bin creation set-up (see Figure~\ref{SD3}). An energy-time entangled state is created after post-selecting the cases where both photons took the short arm or both photons took the long arm of their respective interferometer: for these two cases, both photons arrive at the same time at the detectors A and B.}
\label{SD4}
\end{figure}

\noindent \textbf{Energy-time entanglement} can be seen as the continuous-wave version of time-bin entanglement: by pumping a nonlinear crystal with a continuous-wave laser, two photons can be emitted simultaneously, forming a pair. However, the emission time of this pair is undetermined within the coherence time of the pump laser. This lack of information leads to energy-time entanglement, as first pointed out by Franson~\cite{Franson1989}. A typical set-up used to generate this form of entanglement is sketched in Figure~\ref{SD4}. Note that an essential condition to fulfill in a Franson-type experiment is that $\tau_c, \tau_{det} << \Delta t << \tau_p$, where $\tau_c$ is the coherence time of the emitted photons, $\tau_{det}$ is the time jitter of the detectors, $\Delta t = \Delta L/c$ is the time-difference between the two paths of the interferometer, and $\tau_p$ is the coherence time of the pump laser. In this experiment, a quantum interference occurs since the two processes of both photons having taken the long arm or both photons having taken the short arm are indistinguishable.

\subsection{Physical processes generating photon pairs}
\subsubsection{Parametric processes in nonlinear materials}

\begin{figure}[htb]
\centerline{\includegraphics[width=1\textwidth]{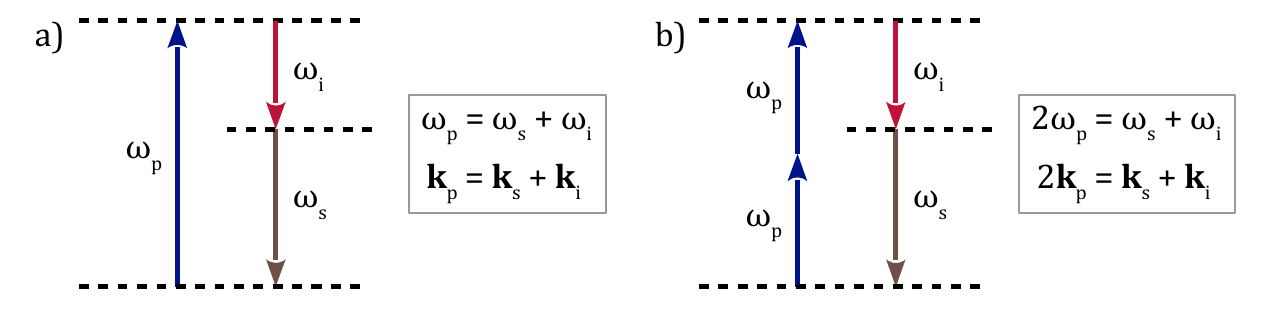}}
\caption{Sketch of the process of parametric down conversion of second a) and third b) order. a) A pump photon with angular frequency $\omega_p$ is annihilated and a pair of photons called signal and idler are created respectively with angular frequencies $\omega_s$ and $\omega_i$. b) Two pump photons with angular frequency $\omega_p$ are annihilated and a pair of signal and idler photons is created. In both conversion processes, the energy and the momentum are conserved.}
\label{SD5}
\end{figure}

Nonlinear optical processes are the most widely used methods to produce entangled photonic quantum states. In a simplified semiclassical model we can express the nonlinear response of a medium to an intense electromagnetic field $\textbf{E}$~\cite{Boyd} by writing the polarization of the material $\textbf{P}$ as $P_i = \epsilon_0\left( \sum_j \chi_{ij}^{(1)} E_j + \sum_{jk} \chi_{ijk}^{(2)} E_jE_k + \sum_{jkl} \chi_{ijkl}^{(3)} E_jE_kE_l \right)$ , where $\chi_{ijk}^{(2)}$  and $\chi_{ijkl}^{(3)}$ are the second and third order nonlinear optical susceptibility tensors with $i,j,k,l = x,y,z$, and the infinite series is truncated at the third order. 
A general result, when including successive terms in the polarization series, is that:
\begin{equation}
\fl
\left|\frac{P_i^{(n+1)}}{P_i^{(n)}}\right| \approx \left|\frac{E}{E_{at}}\right|,
\end{equation}
with $E_{at}$ the characteristic atomic field strength. With a typical electrical field $E \approx 10^5\,V/m$ and with $E_{at} \approx 10^{10}\,V/m$, it results that each further term in the polarization expansion is roughly five orders of magnitude weaker than the previous one. Figure~\ref{SD5} illustrates the process of spontaneous parametric down-conversion of second and third orders.

In the first case a pump photon has a small probability of being converted into a photon pair; this kind of process is called three-wave mixing. In the second case the photon pair is generated from two pump photons; we speak then of four-wave mixing. 
These processes do not involve any transfer of energy between the optical field and the material system, except for a short time interval involving virtual levels (generally of the order of some femtoseconds). In order to have a maximum efficiency in the frequency conversion, both energy and momentum have to be conserved. The conservation of momentum is generally refered to as `phase-matching' because it translates into a condition on the phase velocities of the different interacting waves. Its fulfilment usually requires some refractive index dispersion engineering. The two photons of the down-conversion pair can be produced with the same polarization (type-I process) or orthogonal polarizations (type-II process). The two generated photons, often called signal (s) and idler (i) for historical reasons, can leave the down-converting medium either in the same direction or in different directions, two configurations known as the collinear and non-collinear cases, respectively.

The conservation laws underpinning parametric down-conversion processes create quantum correlations in one or more degrees of freedom describing the state of the photon pair. The main step in the development of practical quantum correlation and quantum entanglement tools has been the development of ultra-bright sources of correlated photons and that of novel principles of entangled states engineering. This also includes entangled states of higher dimensionality and entangled quantum states demonstrating simultaneous entanglement in several pairs of quantum variables (hyper-entanglement). The different techniques that have been developed to achieve efficient nonlinear processes in semiconductor waveguides and produce entangled photon pairs will be discussed in section~4.

\subsubsection{Biexciton-exciton cascade in optically active quantum dots}

\begin{figure}[htb]
\centerline{\includegraphics[width=1\textwidth]{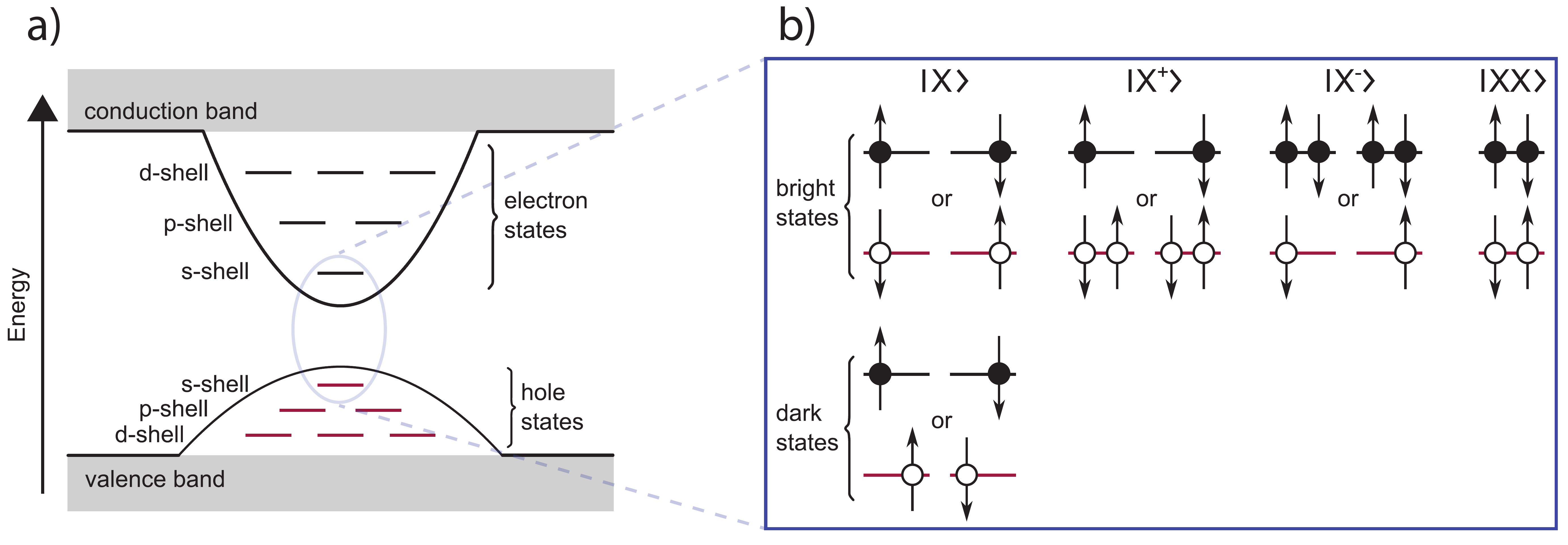}}
\caption{a) Schematic energy potential of a quantum dot. The quantum dot potential can be approximated by a 2D harmonic oscillator model with different effective masses for electron and holes. The first three quantized electronic (black) and hole (red) levels are illustrated. b) Illustration of the nine s-shell states charge carrier configurations. The resulting quantum states are defined by the number of electrons (filled circles) and holes (empty circles) and their respective spin configuration (direction of arrows). The dark exciton state has a parallel spin configuration, resulting in an optically inactive state with a long lifetime.}
\label{KDJ1}
\end{figure}

Optically active semiconductor quantum dots are nanostructures that provide three-dimensional confinement for charge carriers and have a size comparable to the de~Broglie wavelength of the electron~\cite{deBroglie1924}; they are thus called zero-dimensional structures. Typical semiconductor quantum dots consist of a material/alloy with a smaller band gap than the materials surrounding it (also called barrier). If the smaller band gap of the quantum dot lies fully inside the larger band gap of the barrier (straddling gap), this is called type I band alignment. This results in a confining potential for both electrons and holes, which form strongly confined excitons (electron-hole pairs) inside the quantum dot at cryogenic temperatures. The strong confinement leads to the quantization of the particle motion and results in discrete energy levels~\cite{Cibert1986}. This effect is referred to as quantum confinement. The first experimental demonstration of discrete electronic states in zero-dimensional semiconductor nanostructures was done in 1988~\cite{Reed1988}. The labelling of the discrete energy levels follows the convention of atomic physics, giving quantum dots also the name \textit{artificial atoms}~\cite{Kastner1993,Ashoori1996}. A schematic energy potential of a type-I band-aligned quantum dot is shown in Figure~\ref{KDJ1}a. Important parameters for the energetic level structure of quantum dots are not only the size and shape of the quantum dot but also the semiconductor heterostructure composition~\cite{Williamson2000}, the random alloying inside the quantum dot~\cite{Shumway2001,Mlinar2009}, and the surrounding semiconductor barrier. The lowest energetic level in the conduction and valence band is called s-shell. Due to the Pauli exclusion principle, the conduction (valence) band s-shell can be maximally occupied by two electrons (holes) with different spin configuration. This leads to nine different electron-hole pair configurations in the s-shell as shown in Figure~\ref{KDJ1}b. These electron-hole pair configurations are categorized, based on the number of charge carriers, into four quantum states: exciton $\ket{X}$, positively (negatively) charged exciton $\ket{X^+}$ ($\ket{X^-}$), and biexciton $\ket{XX}$. Due to the optical selection rules, two configurations in the s-shell, where both the electron and hole have the same spin configuration are optically inactive and called dark excitons. For a detailed analysis of the dark states the authors suggest to the interested reader the following publications~\cite{Bayer2000,Reischle2008}.

In a simplified picture, one can describe the quantum dot system as a quantum mechanical two-level system, where the excited state is an exciton in the quantum dot and the ground state is an empty quantum dot without any charge carriers. After the excitation of an electron and its relaxation to the s-shell conduction band, the formed exciton recombines and a single photon is emitted~\cite{Michler2000}. In contrast to the probabilistic nature of parametric down-conversion sources, quantum dots can emit on-demand single photons~\cite{He2013,Ding2016,Somaschi2016} when excited resonantly with a oscillatory driving field from the ground state to the excited state (e.g. with a $\pi$-pulse), allowing for coherent control of the two-level system~\cite{Kamada2001}.

Another important excitation state in the s-shell of a quantum dot is that of a fully occupied s-shell with two electrons and two holes, forming two excitons. This is called a biexciton state. If one neglects any non-radiative processes, a direct transition from the biexciton state to the ground state of the quantum dot is not possible with only one single-photon emission. Instead, the biexciton state recombines to one of the two bright exciton states. Note that, due to the Coulomb interaction between the electron and holes, the energy of the biexciton state is different from that of the exciton, where only one electron and hole are confined in the s-shell~\cite{Hu1990}. This difference in emission energy between the exciton photon and the biexciton photon is defined as the biexciton binding energy $E_b$. After the quantum dot s-shell is occupied with two electron-hole pairs, first a single photon from the biexciton state is emitted, leaving the quantum dot in one of the bright exciton states. Then the remaining exciton recombines, emitting another single photon. This is called the biexciton-exciton cascade~\cite{Moreau2001}, where a biexciton single-photon is followed by an exciton single-photon and never the other way around. 
This cascaded photon emission can be used in several schemes to generate entangled photon-pairs from a quantum dot. They will be detailed in Section 4.

\subsection{Indistinguishability of photons -- HOM effect for path entanglement}

\begin{figure}[htb]
\centerline{\includegraphics[width=1\textwidth]{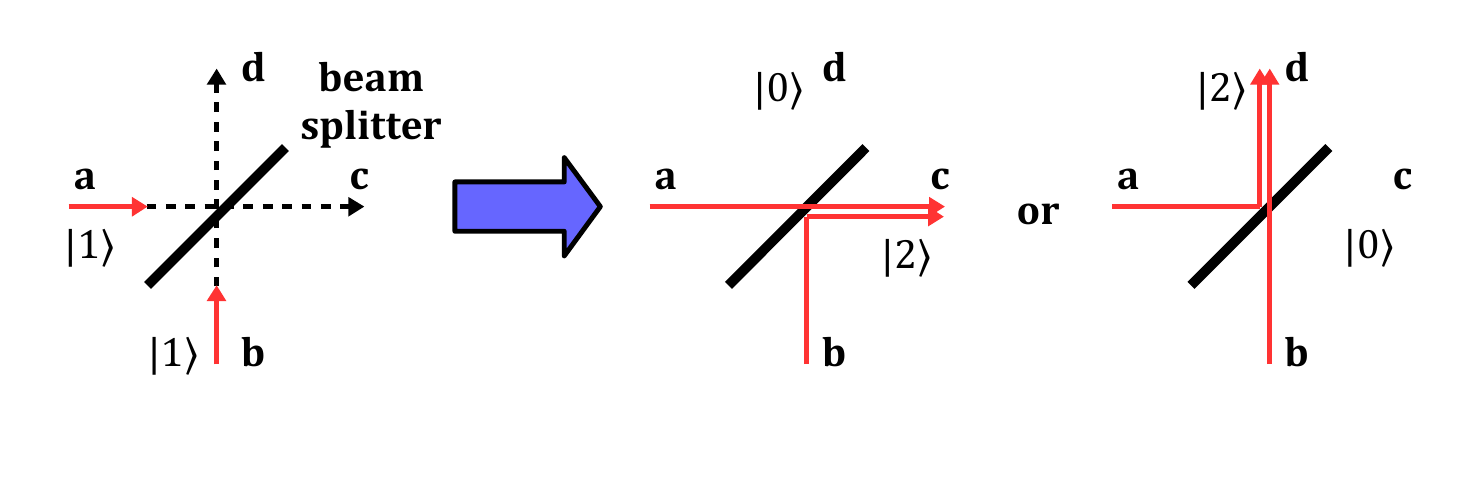}}
\caption{The HOM effect. Indistinguishable photons (red arrows) entering opposite input ports of a 50/50 beam splitter always end up in the same output port, resulting in the entangled state $\frac{1}{\sqrt{2}}(\ket{2}_{c}\ket{0}_{d}+\ket{0}_{c}\ket{2}_{d})$.}
\label{MV1}
\end{figure}

Path entanglement can also be generated by the interference of two indistinguishable photons at a beam splitter.
Usually, when two photons enter a 50/50 beam splitter, each via a different input port, there is a probability of 0.5 that they both end up in the same output port. However, when the two photons are completely indistinguishable, such that it is fundamentally impossible to find out which photon took which entrance, then the photons always take the same output port (Figure~\ref{MV1}). This purely quantum mechanical effect was discovered by Hong, Ou and Mandel~\cite{HOM1987}, and therefore it is known as the Hong-Ou-Mandel (HOM) effect. This effect is based on nonclassical two-photon interference~\cite{Ghosh1986,Ghosh1987}. Photons are indistinguishable when they are, apart from the direction from which they came, the same in all their degrees of freedom (frequency, time, polarization, spatial mode, orbital angular momentum). The HOM effect is thus a method of creating path entanglement. The photons exiting the beam splitter are in the simplest N00N state, the state $\frac{1}{\sqrt{2}}(\ket{2}_{c}\ket{0}_{d}+\ket{0}_{c}\ket{2}_{d})$, where the numbers indicate the number of photons in the output ports $c$ and $d$ of the beam splitter. Two-photon HOM interference forms the basis of entanglement swapping, the creation of entanglement between two particles that never interacted~\cite{Pan1998,Zukowski1993}: this technique is the key ingredient of a quantum repeater. The HOM effect also enables the realization of a linear optical CNOT gate, the basic element of a linear optical quantum computer~\cite{KLM2001}.

Mathematically, the HOM effect can be understood in a simple way~\cite{Pan2012}. We consider two indistinguishable photons entering the 50/50 beam splitter via paths $a$ and $b$ (Figure~\ref{MV1}). The initial state can be written as $\ket{1}_{a}\ket{1}_{b}=a^\dagger b^\dagger\ket{\Omega}$, where $a^\dagger$ and $b^\dagger$ are creation operators and $\ket{\Omega}$ is the vacuum state. The 50/50 beam splitter transforms the modes $a$ and $b$ as $a^\dagger \rightarrow \frac{1}{\sqrt{2}}(t c^\dagger+r d^\dagger)$ and $b^\dagger \rightarrow \frac{1}{\sqrt{2}}(r'c^\dagger+t'd^\dagger)$. The beam splitter transformation can also be written in matrix form~\cite{Zeilinger1981} as 
\begin{equation}
	B=\frac{1}{\sqrt{2}}\begin{pmatrix}
	t & r \\
	r' & t' \\
	\end{pmatrix}.
\end{equation}
For a lossless beam splitter, conservation of probability requires this matrix to be unitary. This unitarity implies that $|r|=|r'| $, $ |t|=|t'| $, and $ r^{*}t'+t^{*}r'=0 $. From these relations one can see that the phase of reflected and transmitted photons must obey $ |r||t| e^{-i(\phi_{r}-\phi_{t'})}+|r||t| e^{i(\phi_{r'}-\phi_{t})}=0$, and therefore $\phi_{r}+\phi_{r'}-\phi_{t}-\phi_{t'}=\pm \pi$. In the case of a symmetric lossless beam splitter, we have $\phi_{r}-\phi_{t}=\phi_{r'}-\phi_{t'}=\frac{\pi}{2}$. Ignoring an overall phase factor, we get $t=t'=1$ and $r=r'=i$. The symmetric lossless beam splitter therefore transforms the input state $a^\dagger b^\dagger\ket{\Omega}$ into $\frac{1}{\sqrt{2}}(c^\dagger+id^\dagger)\frac{1}{\sqrt{2}}(ic^\dagger+d^\dagger)\ket{\Omega}=\frac{i}{2}(c^{\dagger2}+d^{\dagger2})\ket{\Omega}=\frac{i}{\sqrt{2}}(\ket{2}_{c}\ket{0}_{d}+\ket{0}_{c}\ket{2}_{d})$. We see here that the photons take the same exit port of the beam splitter. Because of the indistinguishability of the photons, the terms $\frac{1}{2}c^\dagger d^\dagger\ket{\Omega}$ and $-\frac{1}{2}d^\dagger c^\dagger\ket{\Omega}$ cancel each other. In other words, the two indistinguishable processes where both photons are transmitted and where both photons are reflected destructively interfere. However, when the two photons are distinguishable, for example by different arrival times $t_{1}$ and $t_{2}$, then the terms $\frac{1}{2}c^\dagger(t_{1}) d^\dagger(t_{2})\ket{\Omega}$ and $-\frac{1}{2}d^\dagger(t_{1}) c^\dagger(t_{2})\ket{\Omega}$ do not cancel each other. This relation between HOM interference and degree of indistinguishability has been shown in measurements of the two-photon coincidence rate versus the time delay between the photons~\cite{HOM1987}. Note that HOM interference, where the two-photon coincidence rate goes to 0 for completely indistinguishable particles, is a purely quantum mechanical phenomenon. When interfering classical waves, there can also be a dip in the two-photon coincidence rate, but this dip is never lower than 0.5.

Obtaining a high-visibility HOM interference requires highly indistinguishable photons. For nonlinear semiconductor sources of photon pairs, this is not fundamentally harder to achieve than for more conventional nonlinear sources. If the two photons come from the same pair, very high visibilities can be achieved without much effort, especially for waveguide-based sources. If instead they are heralded from two pairs emitted by two different sources, one also needs to either use a narrow spectral filtering or engineer the spectral emission mode of the sources so as to ensure that the photon pairs are frequency-uncorrelated~\cite{Mosley2008,Walton2004,Pryde2016}.
The Hong-Ou-Mandel effect has been used with several nonlinear semiconductor sources for different purposes. For example, it has allowed to test the indistinguishability of the photons generated by AlGaAs waveguides in two different geometries~\cite{Caillet2010,Autebert2016}, reaching HOM visibilies of $0.85$ and $0.861\pm0.027$ respectively (raw visibilities, i.e. without background noise substraction), for photons from the same pair. In both experiments, the visibility was limited by the high reflectivity of the facets of the waveguides caused by the large difference in refractive index between AlGaAs ($n \approx 3.1$) and air ($n=1$). However, in principle, a near-perfect visibility could be achieved with antireflection coatings. Photon pairs generated from two independent silicon wire waveguide sources have already shown an interference visibility of 0.73 (raw) in a first proof-of-principle experiment~\cite{Harada2011}, thus demonstating their viability for entanglement swapping applications. The HOM effect has also been used to generate path-entanglement from two spiraled silicon waveguide sources with a raw visibility of $0.945\pm0.003$ ($1.000\pm0.004$ net visibility, i.e. with background noise substraction)~\cite{Silverstone2013}.

For photons emitted from quantum dots, very high degrees of indistinguishability have been achieved. The first demonstration of the HOM effect with quantum dot photons was by Santori \textit{et al.}~\cite{Santori2002}, who let single photons from two subsequent excitations of a quantum dot interfere at a beam splitter and measured an indistinguishability of 0.81. By means of resonant excitation of a quantum dot, He \textit{et al.}~\cite{He2013} created indistinguishable photons on demand with an indistinguishability of $0.97\pm0.02$, and Müller \textit{et al.}~\cite{Muller2014} generated indistinguishable entangled photon pairs on demand with indistinguishabilities of $0.86\pm0.03$ for XX photons and $0.71\pm0.04$ for X photons. Wei \textit{et al.}~\cite{Wei2014} and Somaschi \textit{et al.}~\cite{Somaschi2016} optimized the indistinguishability of photons emitted from the same quantum dot a few nanoseconds after each other to $0.995\pm0.007$ and $0.9956\pm0.0045$ respectively. The degree of indistinguishability decreases with increasing time interval between emission events, because of dephasing processes such as charge fluctuations in the vicinity of the quantum dot, but Wang \textit{et al.}~\cite{Wang2016} still achieved an indistinguishability of $0.921\pm0.005$ with a time interval of $14.7\,\mu s$ between the emission events. It is much harder to make two quantum dots emit indistinguishable photons, since the quantum dots consist of a large number of atoms, while the presence or absence of one elementary charge can already have a significant effect on the frequency of the emitted photons. Two-photon HOM interference of single photons from two separate quantum dots was shown by Flagg \textit{et al.}~\cite{Flagg2010}, and Patel \textit{et al.}~\cite{Patel2010}, reaching indistinguishabilities of $0.181\pm0.004$ and $0.33\pm0.01$ respectively, which was recently improved to $0.51 \pm 0.05$
by Reindl et al.~\cite{Reindl2017} using phonon-assisted two-photon resonant excitation. Instead, when using coherently scattered~\cite{Gao2013} or Raman photons~\cite{He2013PRL} from remote QDs the two-photon interference visibilities reported were as high as $0.82 \pm 0.02$ and $0.87 \pm 0.04$, respectively.
One should be aware that various authors use various analyses and correction methods to achieve values for the indistinguishability and comparing those values should be done with care. We gave here the highest reported values and we refer the interested reader to the cited articles for explanations of the analysis and correction methods.

Indistinguishability of photons also opens up other possibilities, apart from the HOM effect, of creating entanglement by quantum interference. For example, Fattal \textit{et al.}~\cite{Fattal2004} created polarization entanglement by giving one of the indistinguishable photons from a quantum dot opposite polarization, interfering both photons at a beam splitter, and postselecting the cases where both photons were detected at opposite output ports.


\section{Measuring entanglement}

As pointed out in the introduction, the correlations between measurement outcomes in quantum mechanical systems puzzled many physicists. In 1935, Einstein, Podolsky and Rosen triggered the debate whether or not the theory of quantum mechanics should be completed with some extra information (hidden variables) able to describe the observed correlations in classical terms~\cite{EPR1935}. A beautiful presentation of the concepts behind the EPR paradox is given in~\cite{Aspect2004}; here we summarize the main ideas discussed in that paper.

\begin{figure}[htb]
\centerline{\includegraphics[width=1\textwidth]{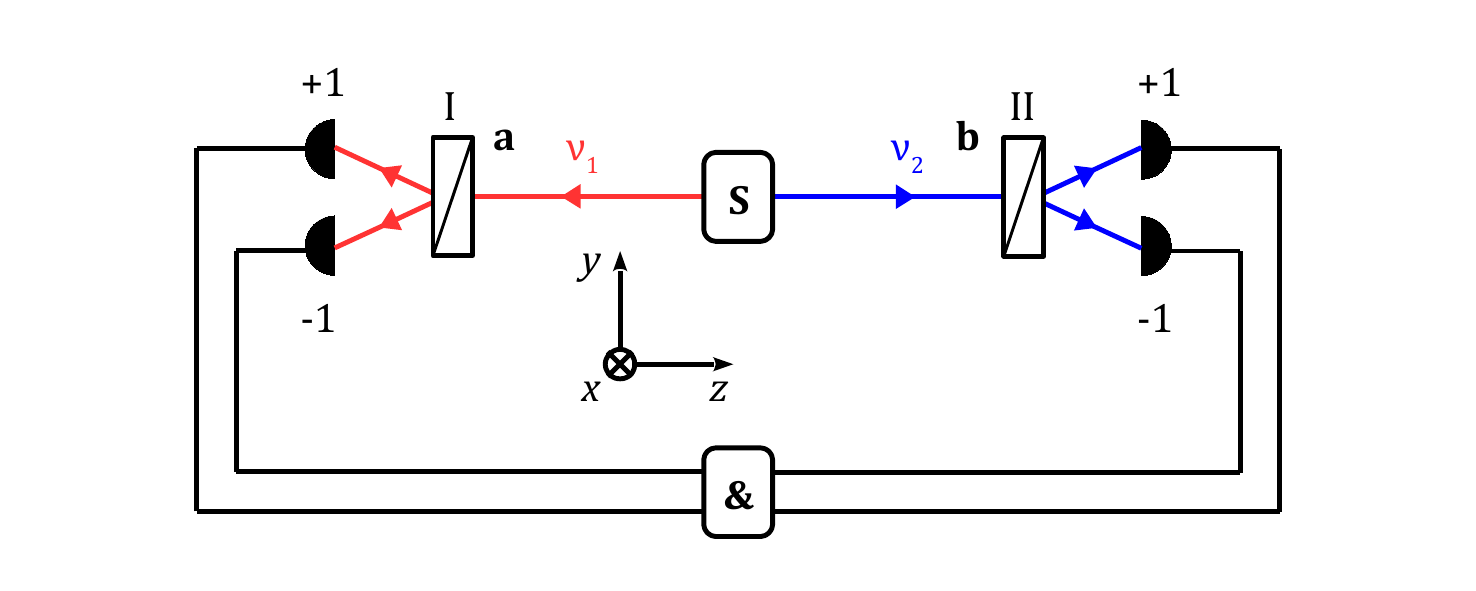}}
\caption{Sketch of the Einstein-Podolski-Rosen-Bohm gedankenexperiment with photons (see the text for details).}
\label{SD6}
\end{figure}

Let us consider the situation sketched in Figure~\ref{SD6}, where a source emits pairs of counterpropagating photons, with respective frequencies $\nu_{1}$ and $\nu_{2}$, that are entangled in polarization. This is an optical variant of Bohm's version of the EPR gedankenexperiment~\cite{Bohm1951}. Let us also suppose, to fix the ideas, that the polarization part of the state vector describing the pair is:
\begin{equation}
\fl
\ket{\Psi}=\frac{1}{\sqrt{2}}\left(\ket{x,x}+\ket{y,y}\right),
\label{Ebell}
\end{equation}
where $\ket{x}$ and $\ket{y}$ are linear polarization states in the coordinate systems reported in Figure~\ref{SD6}. A linear polarization measurement is done using the analysers I and II, having their transmission axis oriented respectively along the directions $a$ and $b$. Each analyser is followed by two detectors, giving the results +1 or -1, according to whether the linear polarization is found parallel or perpendicular to the transmission axis of the analyser. The experimental set-up allows to measure both the single and the joint probabilities of detection (this last operation is indicated with the \& symbol in Figure~\ref{SD6}). It can be easily shown that the single probabilities $P_\pm(a)$ ($P_\pm(b)$) of obtaining results ${\pm1}$ for photon 1 (2) are
\begin{eqnarray}
\fl
P_+(a)=P_-(a)=\frac{1}{2}, \\
\fl
P_+(b)=P_-(b)=\frac{1}{2}.
\end{eqnarray}
As usual, quantum mechanics does not tell us the measurement outcomes: it only gives the probabilities of measurement outcomes. The probabilities $P_{\pm\pm}(a,b)$ of joint detections are
\begin{eqnarray}
\fl
P_{++}(a,b)=P_{--}(a,b)=\frac{1}{2}\cos^2(\widehat{a,b}), \\
\fl
P_{+-}(a,b)=P_{-+}(a,b)=\frac{1}{2}\sin^2(\widehat{a,b}),
\end{eqnarray}
where $\widehat{a,b}$ is the angle between directions $a$ and $b$.\\
If we consider the particular case where the analysers are parallel ($\widehat{a,b}=0$), the joint detection probabilities are
\begin{eqnarray}
\fl
P_{++}(a,b)=P_{--}(a,b)=\frac{1}{2}, \\
\fl
P_{+-}(a,b)=P_{-+}(a,b)=0,
\end{eqnarray}
which shows that there is a total correlation between the results of polarization measurement of the two photons of each pair.\\
To quantify the amount of correlations between random events, we introduce the correlation coefficient, which is equal to
\begin{equation}
\fl
E(a,b)=P_{++}(a,b)+P_{--}(a,b)-P_{-+}(a,b)-P_{+-}(a,b).
\label{EvsP}
\end{equation}
Using the expressions for the joint detections probabilities written above, we find
\begin{equation}
\fl
E(a,b)=\cos(2\widehat{a,b}),
\label{Ecos}
\end{equation}
which, in the case of parallel analysers, gives $E=1$.\\
When we perform a polarization measurement on photon 1, then we know with certainty also the outcome of the polarization measurement on photon 2, if the two analysers have been placed in parallel. Einstein, Podolsky, and Rosen (EPR) assumed that the two measurements took place at a large distance from each other, so that the measurement on photon 1 could  not affect the outcome of the measurement on photon 2. Since one can with certainty predict the outcome of the measurement on photon 2, according to the reasoning of EPR, photon 2 must have had this measured polarization already before the measurement took place. In other words, there must have been an element of physical reality that determined the outcome of the polarization measurement on photon 2. The same argument applies to photon 1 and to all parallel polarization angles, even to variables that are non-commutative in quantum mechanics. These polarization properties that would exist prior to measurement and determine the measurement outcomes are not contained in quantum mechanics and are therefore referred to as `hidden variables'. In the view of EPR, a complete theory of nature should include these hidden variables.

The assumption of locality, namely that the setting of the analyser and the measurement outcome on one side cannot influence the measurement outcome on the other side, is considered to be a very natural restriction for any otherwise most general hidden-variable (or `realistic') model. In 1964, Bell translated into mathematics the consequences of the `hidden variable' theories; he found that the correlation between the two measurements, as predicted by any such local realistic model, must necessarily comply with a set of inequalities. The most widely used form reads~\cite{CHSH}:
\begin{equation}
\fl
S(a,b,a',b') = |E(a,b) - E(a,b') + E(a',b) + E(a',b')| \leq 2,
\end{equation}
where $E(a, b)$ is the correlation coefficient of measurements along $a$, $a’$ and $b$, $b’$, defined in Equation~\ref{EvsP}. $S$ is called the Bell parameter and has the meaning of a second-order correlation. From Equation~\ref{Ecos}, it can easily be seen that the Bell state of Equation~\ref{Ebell} violates this inequality with $S=2\sqrt{2} \approx 2.828$ for a specific set of analyser directions: $a=0^{\circ}$, $a'=45^{\circ}$, $b=22.5^{\circ}$ and $b'=67.5^{\circ}$. This value for $S$ is the maximal value that a quantum mechanical state can achieve. The conclusion is that a system that can be described by a local realistic theory cannot mimic the behavior of entangled states and, hence, that quantum theory must be a non-local or a non-realistic one. In other words, if hidden variables determine the measurement outcomes, then non-local interactions must exist. Non-local interactions here can be defined as interactions that exceed the speed of light. Since the first experiments in the ’70s and ’80s, the violation of Bell inequalities has become a widely spread protocol, even if closing all significant loopholes in its experimental implementation is still very challenging and has been only recently achieved~\cite{Hensen2015,Giustina2015,Shalm2015,Rosenfeldarxiv} as mentioned in the introduction. Today the violation of Bell inequalities is a standard test to demonstrate the ability of a source to generate strongly entangled states (whenever $S>2$). One of the most promising applications of photonic states that violate Bell inequalities is the so-called device-independent quantum key distribution (DI-QKD) which is a quantum cryptographic protocol that relies on non-locality (or non-realism) and whose security does not depend of assumptions about the physical devices used to implement the protocol. The idea was first introduced in~\cite{Ekert1991} and has been an active research subject for about ten years~\cite{Acin2006,Pironio2009}.

Bell inequalities are not violated by all entangled states, so this test can only be used as a witness of a certain type of entanglement. The most complete approach to measure a quantum state is the reconstruction of its density matrix, which is called quantum tomography. This has been investigated for the first time in~\cite{White1999} for photon pairs entangled in polarization, but it can be generalized to any degrees of freedom. The measurement scheme is the same as in Figure~\ref{SD6}. In order to do a complete tomography, there are just more analyser directions to measure than when testing whether the two-particle system could be correctly described by a local realistic theory. The density matrix is then reconstructed from the statistical outcomes of different joint projection measurements. One of the fundamental references describing in detail the theory underpinning the reconstruction of a density matrix starting form experimental data is~\cite{James2001}. The authors discuss a tomographic measurement method and a maximum likelihood technique requiring a numerical optimization but allowing to reconstruct physical density matrices (the codes to perform a quantum state tomography based on this method are available on the website of P. Kwiat’s group, http://research.physics.illinois.edu/QI/Photonics/Tomography/).
The density matrix is a Hermitian matrix with unitary trace which gives a complete description of an arbitrary quantum state. In the case of a pure state $\ket{\psi}$  (which can be a coherent superposition of pure states), it is defined as $\rho_\psi = \ket{\psi}\bra{\psi}$; it is thus a projection operator: $\Tr(\rho_\psi^2)=1$. In the case of a statistical mixture of $n$ qubits, it is written as $\rho_{mix}=\sum_{i=1}^{2^n} P_i\ket{\psi_i}\bra{\psi_i}$, with $\sum_{i=1}^{2^n} P_i = 1$ and $\langle\psi_i\ket{\psi_j}=\delta_{ij}$. $\Tr(\rho_{mix}^2) < 1$ when more than one of the $P_i$ is different from zero.
The density matrix corresponding to a coherent superposition of pure states is characterized by the presence of off-diagonal terms, while in the matrix corresponding to a statistical mixture these terms are zero.

From the density matrix of a two-qubit state, one can extract several indicators giving information on the nature of the state the matrix represents~\cite{Altepeter2005}.\\
A first indicator is the fidelity~\cite{Jozsa1994}, which is a measure of the overlap between the reconstructed state $\rho$ and a known state $\rho_0$: $F_{\rho_0}(\rho) = \left[\textrm{Tr}\left(\sqrt{\sqrt{\rho_0}\rho\sqrt{\rho_0}}\right)\right]^2$. If $\rho_0$ can be written as a pure state $\rho_0 = \ket{\psi_0}\bra{\psi_0}$, the expression of the fidelity reduces to $F_{\ket{\psi_0}}(\rho) = \textrm{Tr}\left(\ket{\psi_0}\bra{\psi_0}\rho\right)$. The fidelity is equal to 1 if the states are perfectly identical, it is equal to zero if they are orthogonal. The fidelity between the reconstructed state and a Bell state is an entanglement witness: if this Bell state fidelity is larger than 1/2 then the state is entangled~\cite{White2007}.\\
Another indicator is the entropy defined either as the Von Neumann entropy $S_{VN}(\rho)=-\textrm{Tr}\left(\rho\,\textrm{log}_2\,\rho\right)$ or as the linear entropy $S_L(\rho)=\frac{4}{3}(1-\textrm{Tr}\left(\rho^2\right))$, where $\textrm{Tr}\left(\rho^2\right)$ is the purity of the state $\rho$. The entropy quantifies the mixedness of the reconstructed state: the linear entropy $S_L$ is equal to 0 for a pure state and to 1 for a maximally mixed state. For a given value of entropy, there is an upper bound to the amount of entanglement that may be present in the state~\cite{Wei2003}.\\
To quantify the amount of entanglement present in a two-qubit state, several entanglement measures have been introduced~\cite{Plenio2007}. The most widely used by experimentalists are:
\begin{itemize}
\item[-] the concurrence~\cite{Wootters1998}: $C(\rho) = \max{\{0;\sqrt{r_1}-\sqrt{r_2}-\sqrt{r_3}-\sqrt{r_4}\}}$, where $r_1 \geq r_2 \geq r_3 \geq r_4$ are the eigenvalues of $\rho(\sigma_y\otimes\sigma_y)\rho^*(\sigma_y\otimes\sigma_y)$, where $\rho^*$ is the element-wise complex conjugate of $\rho$ and $\sigma_y=i\ket{0}\bra{1}-i\ket{1}\bra{0}$ is a Pauli matrix.
\item[-] the tangle~\cite{Wootters2000}: $T(\rho) = C^2(\rho)$.
\item[-] the negativity~\cite{Vidal2002}: $N(\rho) = \frac{||\rho^{T_A}||-1}{2}$, where $\rho^{T_A}$ is the partial transpose of $\rho$ and $||...||$ is the trace norm. The negativity is directly related to the Peres criterion (or PPT criterion) for entanglement~\cite{Peres1996} which says that a density matrix presents at least one negative eigenvalue under partial transposition if and only if it represents an entangled state.
\end{itemize}
All of the above three measures reach a minimum value of 0 for separable states and a maximal value of 1 for maximally entangled states, such as Bell states.


\section{Solid-state devices generating entangled photons}

\subsection{AlGaAs-based sources using three-wave mixing}

GaAs and related compounds are serious candidates for miniaturizing and integrating different quantum components in the same chip for generation, manipulation and detection of quantum states of light. Indeed, this platform combines a large second order optical susceptibility~\cite{Ohashi1993} with well-mastered growth and processing techniques, as well as a direct band-gap. It also allows a monolithic integration of entangled photon sources with superconducting nanowire single-photon detectors (SNSPD) through gallium arsenide (GaAs) waveguides~\cite{Sprengers2011,Pernice2012}.
However, since the AlGaAs compounds possess a zinc-blende (cubic) structure, they have isotropic linear optical properties and therefore lack natural birefringence. For this reason, efficient second-order parametric down-conversion requires alternative strategies of phase-matching, which will be rapidly reviewed in the following section. An extended review on phase-matching techniques in AlGaAs waveguides can be found in~\cite{Helmy2010}.

\subsubsection{Phase-matching techniques}

\begin{figure}[htb]
\centerline{\includegraphics[width=1\textwidth]{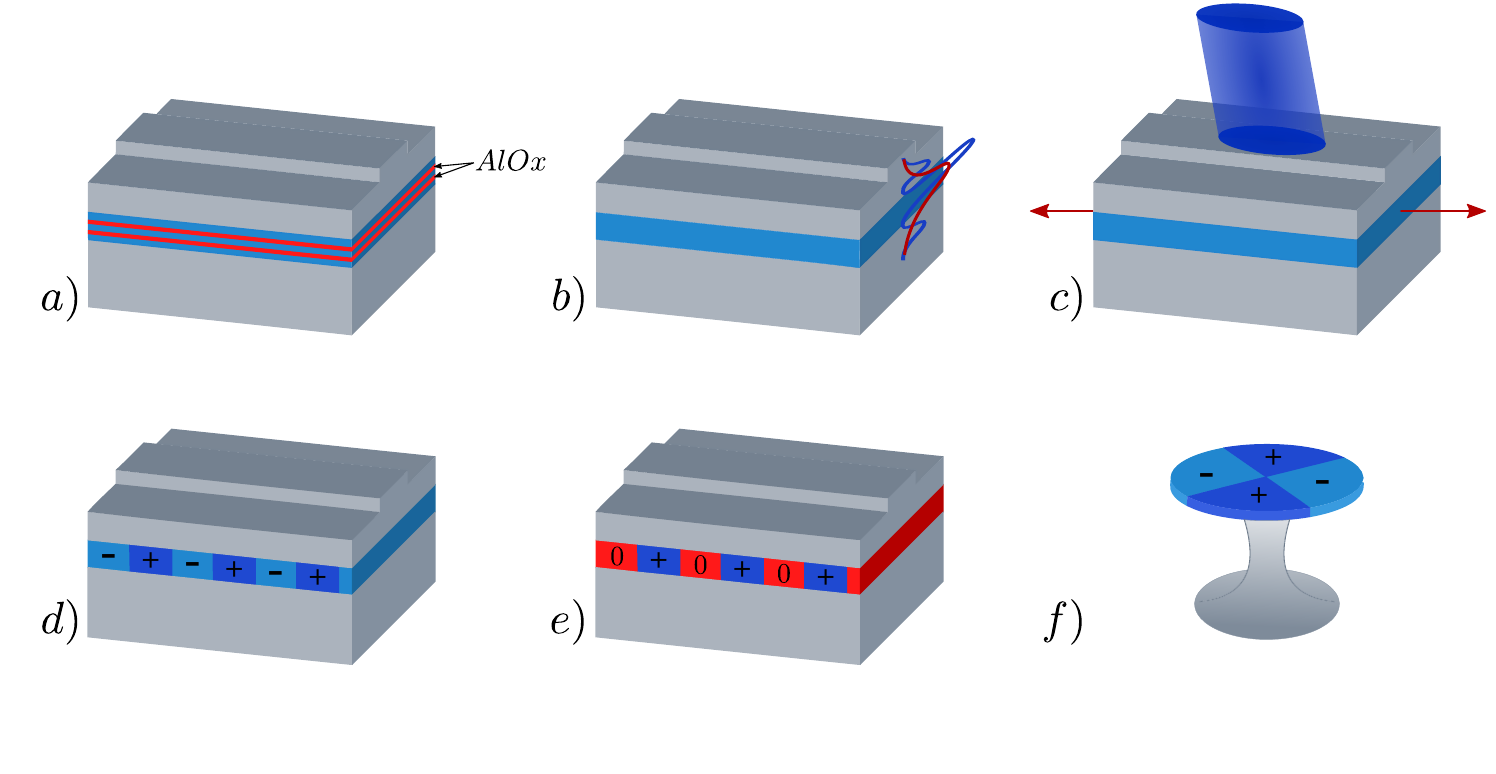}}
\caption{Phase-matching schemes implemented in AlGaAs waveguides (see the text for a detailed description): a) form birefringence phase-matching; b) modal phase-matching; c) counterpropagating phase-matching; d) domain-reversal quasi-phase-matching; e) domain-disordered quasi-phase-matching; f) quasi-phase-matching in circular structures. (\copyright C. Autebert)}
\label{SD7}
\end{figure}

\begin{figure}[htb]
\centerline{\includegraphics[width=1\textwidth]{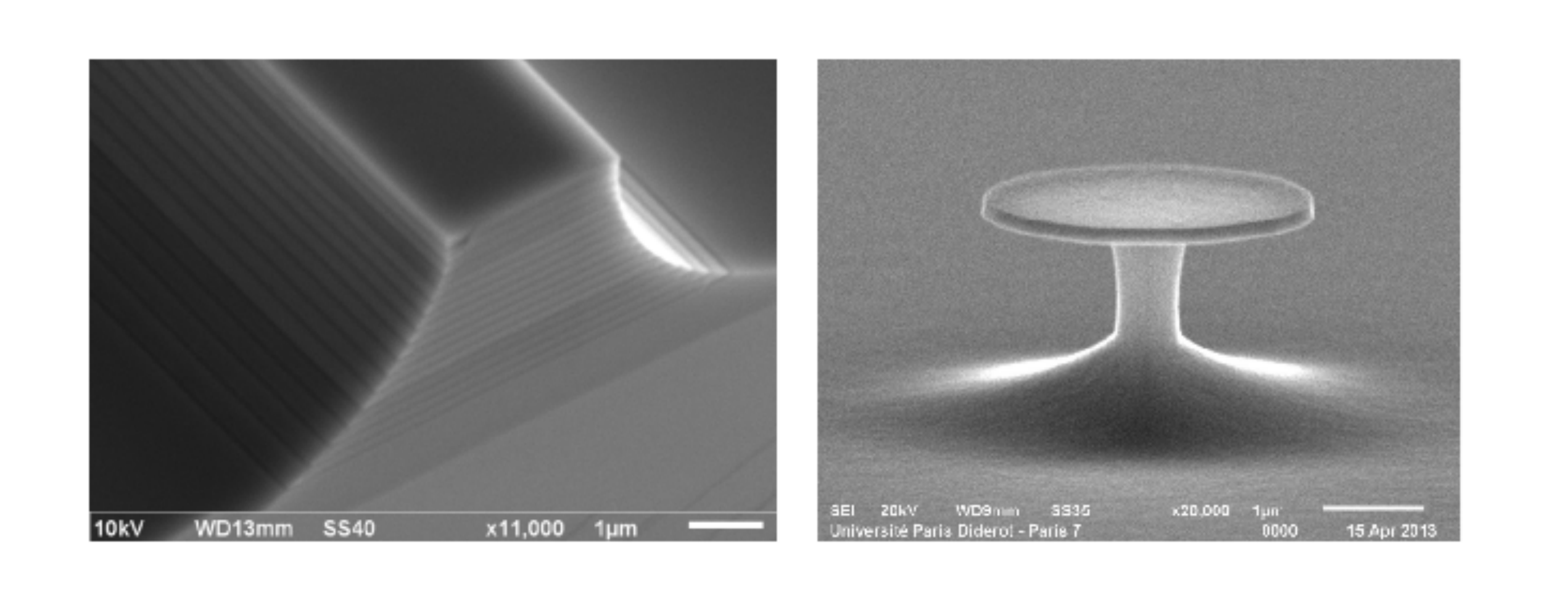}}
\caption{Scanning electron microscope picture of an AlGaAs ridge waveguide designed for modal phase-matching (left) and of an AlGaAs disk suspended over a GaAs pedestal confining whispering gallery modes for quasi-phase-matching (right). (\copyright Laboratoire Matériaux et Phénomènes Quantiques, Université Paris Diderot)}
\label{SD8}
\end{figure}

Epitaxial layers of AlGaAs compounds are usually grown on GaAs substrates along the $<$001$>$ plane and are cleaved along $<$110$>$ planes.
These materials, in bulk layers, present one independent second-order tensor element $\chi_{xyz}^{(2)}$ and, overall, six elements (obtained by interchanging the coordinates with all possible permutations). Additional tensor elements can be obtained by using quantum confined heterostructures to break the crystal symmetry~\cite{Rosencher1991}. However, in the visible and near infrared ranges, where photon-pair sources for quantum information are needed, the effect is too small to be useful.
In straight AlGaAs waveguides, the waves participating to the nonlinear process are usually guided along the $<$110$>$ direction; this implies that the modes have either a transverse-electric (TE) polarization along $<$110$>$ or a transverse-magnetic (TM) polarization along $<$001$>$.
In these last two decades, numerous techniques have been investigated to achieve efficient second-order frequency conversion in AlGaAs devices: the employed phase-matching strategies can be grouped as either exact phase-matching or quasi-phase-matching. Figure~\ref{SD7} presents a sketch of the different phase-matching techniques existing in AlGaAs waveguides and Figure~\ref{SD8} shows scanning electron microscope images of two actual devices.

Among the exact phase-matching techniques, one of the first that have been demonstrated consists in engineering a multilayered structure to provide an artificial form-birefringence phase-matching (FBPM) (Figure~\ref{SD7}a). In this scheme, the artificial birefringence that compensates for the dispersion between the interacting modes is induced by a sub-wavelength refractive-index modulation along the vertical direction, provided by a stack of alternating high-index AlGaAs and low-index aluminum oxide (AlOx) layers. The latter are obtained from Al$_{0.98}$Ga$_{0.02}$As layers by a lateral selective wet oxidation ocuring after etching. Since the first demonstration of AlAs oxidation~\cite{Dallesasse1990}, such a process has progressed a lot in the last two decades, leading to reliable parameters for the kinetics of the reaction, and having recently led to the demonstration of the first integrated optical parametric oscillator (OPO) around $2\,\mu m$~\cite{Savanier2013}. This technique is attractive because no oxide deposition or regrowth is needed; moreover the high overlap of the interacting waves (which are all fundamental modes) allows reaching high frequency conversion efficiencies~\cite{Savanier2011a}. However two main issues have still to be solved: first, linear optical losses are significant both for pump beams around $775\,nm$ and twin photons around $1550\,nm$. This is due to the oxidation process, which induces the formation of absorption centers and of roughness at the interfaces between the AlOx and the AlGaAs layers~\cite{Savanier2011b}. Second, the integration of active devices is hindered by the insulating nature of AlOx, which would constitute a barrier to electrical current injection.

An alternative exact phase-matching scheme is modal phase-matching (MPM) (Figure~\ref{SD7}b and~\ref{SD8}a): in this case, the phase-velocity mismatch is compensated by a multimode waveguide dispersion engineering. Although the overlap integral of the modes involved in the nonlinear process is in general smaller than in the FBPM scheme, no oxidation process is required for MPM. An important advantage of this solution is therefore its compatibility with electrical injection, thus allowing the integration of a laser action with nonlinear effects. The interacting modes can either be guided by total internal reflection (TIR)~\cite{Moutzouris2003,Ducci2004} or by a photonic bandgap effect, such as Bragg reflection (BR)~\cite{Yeh1976,Helmy2006}. This second approach has proved to be more interesting since modes that are confined by Bragg reflectors can have an effective index much lower than the material indices of the waveguide constituents, which gives more flexibility for device engineering. In particular, it helps avoiding ageing problems via the reduction of the total aluminum content.

The alternative to exact phase-matching is quasi-phase-matching (QPM) (Figure~\ref{SD7}d and e). In this case, the phase-mismatch accumulated by the interacting waves during propagation is periodically corrected along the length of the waveguide, by modulating the nonlinear susceptibility at intervals correponding to the coherence length $L_c$. Two main fabrication techniques have been demonstrated up to now: in the domain-reversal QPM technique (Figure~\ref{SD7}d), the nonlinear tensor $\chi^{(2)}$ is alternated in sign from positive to negative; while in the domain-suppressed QPM (Figure~\ref{SD7}e), the magnitude of $\chi^{(2)}$ is periodically suppressed such that it alternates between regions of high nonlinearity and low nonlinearity. 
The domain-reversal technique has been demonstrated by periodically rotating the orientation of the crystal by 90$^{\circ}$ about the $<$001$>$ crystal axis, either by wafer bonding~\cite{Yoo1995,Yoo1996} or by orientation-patterned regrowth~\cite{Skauli2002}, this latter technique having led to the demonstation of optical parametric oscillators~\cite{Vodopyanov2004}.
Domain suppression has been achieved by altering the material composition along the waveguide, for example by etching a grating and replacing the removed material with one having a different value of $\chi^{(2)}$. This fabrication method, technologically simpler, allows to achieve smoother domain interfaces~\cite{Rafailov2001}. 
Numerous challenges have still to be faced with the above mentioned QPM methods since the devices demonstrated up to now suffer from quite high scattering losses. For this reason, research studies aiming at finding solutions based on a more simple fabrication process are very active. The most recent achievements in this field have been done on a post-wafer-growth process known as quantum well intermixing (QWI) and on the utilization of QPM in disk resonators.
In QWI, a quantum well structure is used as the core of the waveguide, then one or several processes are used to introduce point defects into the semiconductor crystal lattice, which promote the lattice atoms diffusion under a rapid thermal annealing~\cite{Helmy2010}. This diffusion process modifies the optical properties of the device, such as the absorption/emission bands, the refractive index and the nonlinearity. QPM gratings can be formed by periodically intermixing the quantum well structure to suppress $\chi^{(2)}$~\cite{Aitchison1998}. The advantage of this approach is that it does not require etch-and-regrowth processes, it should thus allow keeping the scattering losses low.
Different types of cores have been tested, including asymmetric quantum wells~\cite{Bouchard2000}, asymmetric coupled quantum wells~\cite{Aitchison1998} and GaAs/AlGaAs super-lattices. This last solution has been shown to be the most efficient one; second harmonic generation has been reported by using QPM gratings either realized by impurity-free vacancy disordering~\cite{Helmy2000} or ion implantation-induced disordering~\cite{Zeaiter2003}. Recent improvements in the super-lattice design and in the ion implantation-induced technique have led to the demonstration of photon-pair generation by SPDC under CW pumping~\cite{Sarrafi2013}.

An alternative possibility to achieve QPM with AlGaAs exploits its $\bar{4}$ crystal symmetry which can exhibit an effective $\chi^{(2)}$ modulation when the fields propagate in curved geometries (such as in microrings or microdisks) (Figure~\ref{SD7}f and~\ref{SD8}b). A 90$^\circ$ rotation about the $<$001$>$ axis is equivalent to a crystallographic inversion, and hence fields propagating around the $<$001$>$ axis in an uniform AlGaAs microdisk effectively encounter four domain inversions per roundtrip. Thus, the $\bar{4}$ crystal symmetry allows QPM to be achieved without externally produced domain inversions. Following this approach, second harmonic generation has been reported in GaAs~\cite{Kuo2014} and AlGaAs~\cite{Mariani2014} suspended microdisks. The high quality factors of these resonators result in a strong field enhancement inside the cavity, combing efficient frequency conversion and small footprint devices; research is under way to demonstrate SPDC.

A last phase-matching geometry that has been exploited to achieve SPDC in AlGaAs waveguides is based on a transverse pump configuration~\cite{Lanco2006,Orieux2011} (Figure~\ref{SD7}c). In this case, a pump field impinging on top of a multilayer AlGaAs ridge waveguide generates two counterpropagating, orthogonally polarized and waveguided twin photons. Momentum conservation on the propagation axis is satisfied by the counterpropagation of the two down-converted photons, while in the epitaxial direction it is satisfied by alternating AlGaAs layers with different aluminum concentrations (having nonlinear coefficients as different as possible) to implement a quasi-phase-matching (QPM) scheme. As a consequence of the opposite propagation directions of the generated photons, two type-II phase-matched processes can occur simultaneously: the first one where the signal (s) photon is TE polarized and the idler (i) photon is TM polarized, and the second one where the signal photon is TM and the idler one is TE. 

\subsubsection{Entanglement generation}

The efforts presented in the previous section to achieve efficient phase-matching have led to the demonstration of a large number of AlGaAs-based parametric sources emitting in different spectral ranges for various applications. The generation of entangled photons has been achieved with three different phase-matching geometries: counterpropagating phase-matching, modal phase-matching with Bragg reflectors and quasi-phase-matching with QWI.

\begin{figure}[htb]
\centerline{\includegraphics[width=1\textwidth]{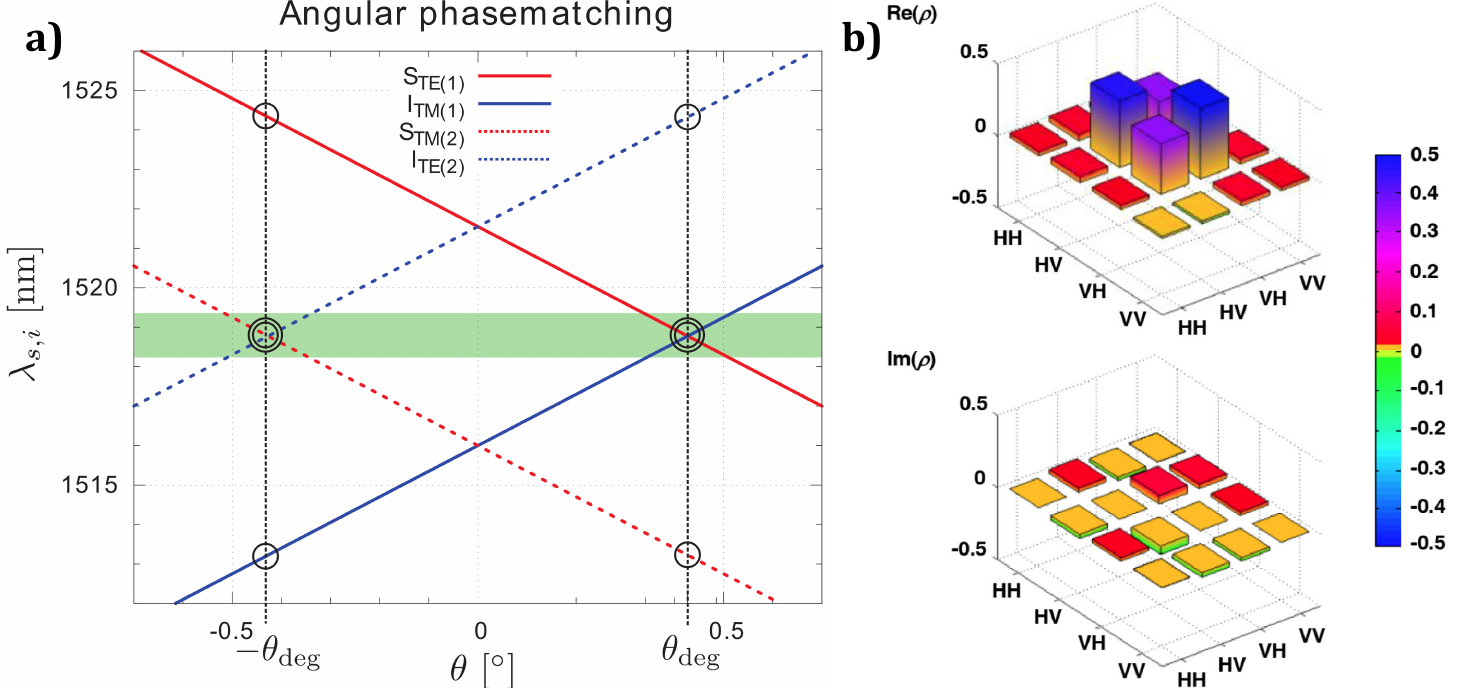}}
\caption{a) Tunability curves of the source of entangled photon pairs based on the counterpropagating phase-matching scheme. This graph shows, for each possible angle of incidence $\theta$ of the pump beam, the wavelength $\lambda_{s,i}$ of the emitted signal (red lines) and idler (blue lines) photons for both phase-matched processes (1) (full lines) and (2) (dotted lines). b) Real and imaginary parts of the tomographically reconstructed density matrix $\rho$ of the two-photon state generated by pumping the source at both angles $\pm \theta_{deg})$ (green area on panel a)). (\copyright 2013 APS, reproduced with permission from~\cite{Orieux2013})}
\label{SD9}
\end{figure}

The first demonstration has been obtained with the counterpropagating phase-matching scheme~\cite{Orieux2013}. In this geometry, the existence of two simultaneously phase-matched processes allows to directly generate Bell states, as illustrated by the tunability curves shown in Figure~\ref{SD9}a. This graph shows that by simultaneously pumping the device at the specific angles $+\theta_{deg}$ and $-\theta_{deg}$ it is posssible to generate two photons (from either process) with identical central frequencies. By filtering out the residual nondegenerate photons that can also be emitted, in principle, a maximally polarization-entangled two-photon state can be obtained. The two-photon state was experimentally generated and then analysed by quantum tomography. The measured density matrix, reported in Figure~\ref{SD9}b, shows a raw fidelity (without background noise substraction) of $0.83\pm0.04$ to the Bell state $\ket{\Psi^+}$ and a concurrence $C$ of $0.68 \pm 0.07$. A theoretical model, taking into account the spatial profile of the pump beam, has been developped to understand how to control the amount of entanglement generated with this source~\cite{Boucher2014}.\\
This geometry offers a particular versatility in the generated quantum state, since both the spatial and the spectral mode shaping of the pump beam allows tailoring the frequency correlations between the two photons of each pair~\cite{Walton2004}. Recently, a technique based on difference frequency generation, has led to the high-resolution reconstruction of the joint spectrum of two-photon states~\cite{Eckstein2014}. This streamlined technique should help in the engineering of the quantum light states at a significantly higher level of spectral detail than previous techniques, enabling future quantum optical applications based on time-energy photon correlations~\cite{Boucher2015}.

Despite the interesting features of the counterpropagating phase-matching scheme, a full integration of the pump laser with the nonlinear medium remains challenging and has not been demonstrated yet. With the aim of developing electrically driven on-chip sources, several teams are working on devices based on modal phase-matching instead. In a first generation of sources, the interacting modes were guided by total internal reflection~\cite{DeRossi2004} but the high aluminium content that was required in these heterostructures to achieve the phase-matching led to problems of oxidation and rapid ageing. More recently, the utilization of Bragg reflectors has added more flexibility to the modal engineering, allowing to demonstrate the first electrically driven photon-pair source operating at room temperature~\cite{Boitier2014}. Up to now, the generation of entanglement with this type of phase-matching has been demonstrated only in passive devices (i.e. with an external pump laser). In~\cite{Valles2013} and~\cite{Horn2013} the authors have demonstrated polarization entanglement by exploiting a type-II process. In the first case, a Bell-type experiment has been performed and a Bell inequality has been violated with a maximum value for the parameter $S$ of $2.61 \pm 0.16$. In the second case, a complete measurement of the density matrix has been done, leading to a value for the concurrence $C$ of 0.52. More recently, energy-time entanglement has also been tested through a Franson experiment~\cite{Autebert2016}, leading to the demonstration of a source able to simultaneously emit indistinguishable and entangled pairs of photons.

Finally, the recent advances in the fabrication of quasi-phase-matched waveguides by QWI have led to the reduction of the level of optical losses, that were formerly a strong hindrance for this approach. In~\cite{Sarrafi2015}, the generation of energy-time entangled photons generated by a type-I process in such waveguides has been obtained, with a Bell-parameter $S$ of $2.687 \pm 0.013$.

An overview of the two-photon entanglement results, reported above, in AlGaAs nonlinear devices is given in Table~\ref{table1}.

\begin{table*}[htbp]
\begin{center}
\begin{tabular}{ | c | c | c | c | c | }
\hline
\multirow{2}{*}{Ref.} & Phase-matching & Entanglement & Test & Result \\
 & type & type & & \\
\hline\hline
\multirow{2}{*}{\cite{Orieux2013}} & counterpropagating & polarization & concurrence & $C = 0.68 \pm 0.07$ \\
 & & & Bell-state fidelity& $F_{\ket{\Psi^+}} = 0.83 \pm 0.04$ \\
\hline
\multirow{2}{*}{\cite{Horn2013}} & modal & polarization & concurrence & $C = 0.52$ \\
 & & & Bell-state fidelity& $F_{\ket{\Psi^+}} = 0.83$ \\
\hline
\cite{Valles2013} & modal & polarization & Bell inequality & $S = 2.61 \pm 0.16$ \\
\hline
\cite{Autebert2016} & modal & energy-time & Bell inequality & $S = 2.70 \pm 0.10$ \\
\hline
\cite{Sarrafi2015} & QPM (QWI) & energy-time & Bell inequality & $S = 2.687 \pm 0.013$ \\
\hline
\end{tabular}
\caption{Comparison between integrated AlGaAs entangled photon-pair sources. The numbers given in the column ``Results'' all correspond to raw values (i.e. without any background noise substraction). Note that a quantitative comparison between these different results is not straightforward as the experimental conditions were not the same (in particular the single-photon detectors and the spectral filters were different).}
\label{table1}
\end{center}
\end{table*}

\subsection{Silicon-based sources using four-wave mixing}

Crystalline silicon has a cubic crystalline structure and does not present second-order optical nonlinearities. However, it exhibits strong third-order nonlinearities, including Kerr effect and Raman gain, allowing nonlinear interactions at relatively low power levels in Silicon-On-Insulator (SOI) waveguides. Several phenomena are investigated for their applications in telecommunications, including stimulated Raman scattering, self- and cross-phase modulation and four-wave mixing. The main problem encountered with Si waveguides is two-photon absorption, a nonlinear process occurring when the energy of the propagating photons exceeds half the bandgap energy $E_g$ ($E_g = 1.1\,eV$ corresponds to a wavelength of about $1.1\,\mu m$ and $\frac{1}{2}E_g$ to about $2.3\,\mu m$). A useful figure of merit can be defined to compare different materials: $F_n=n_2/(\lambda\beta)$, where $n_2$ is the Kerr coefficient (the real part of the nonlinear index) and $\beta$ is the two-photon absorption coefficient (the imaginary part of the nonlinear index). Table~\ref{table2} gives a comparison of different materials of actual interest for integrated photonics~\cite{Grassani2013}.

\begin{table*}[htbp]
\begin{center}
\begin{tabular}{ | c | c | c | c | }
\hline
Material & $n_2$ (cm$^2$/W) & $\beta$ (cm/GW) & $F_n$ \\
\hline\hline
Si & $4.5\times10^{-14}$ & 0.8 & 0.37\\
\hline
SiO$_2$ & $2.2\times10^{-16}$ & - & -\\
\hline
GaAs & $15.9\times10^{-14}$ & 10.2 & 0.10\\
\hline
AlGaAs & $15\times10^{-14}$ & 0.5 & 2\\
\hline
As$_2$S$_3$ & $2.9\times10^{-14}$ & $<$ 0.001 & $>$ 193\\
\hline
As$_2$Se$_3$ & $12\times10^{-14}$ & 0.1 & 8\\
\hline
Ge & $37.9\times10^{-14}$ & 1500 & $\approx$ 0.001\\
\hline
\end{tabular}
\caption{Comparison of third-order nonlinear optical coefficients at $\lambda = 1.5\,\mu m$ for different materials. Silica is practically not affected by two-photon absorption in the near infrared, because of its large band gap of $8.9\,eV$. The values of $n_2$ and $\beta$ reported for Germanium are given at $\lambda = 3\,\mu m$ and $\lambda = 2\,\mu m$ respectively. From~\cite{Grassani2013}.}
\label{table2}
\end{center}
\end{table*}

Reference~\cite{Lin2006} reports the first theoretical study demonstrating the interest of producing photon pairs through spontaneous four-wave mixing (SFWM) in silicon waveguides. In this paper, the authors present a theory for quantifying the quality of generated photon pairs, showing that they not only exhibit high correlation qualities because of the absence of spontaneous Raman scattering (SpRS) but also have a high spectral brightness that is comparable with all other photon-pair sources. Indeed, one of the strong advantages of silicon with respect to dispersion-shifted fibres is that, since the Raman spectrum of monocrystalline silicon is $15.6\,THz$ away from the pump frequency, with a width of about $100\,GHz$, the SpRS photons in silicon can be significantly suppressed by setting the signal/idler frequencies away from the Raman peak. Thanks to the maturity achieved by silicon technology, the generation of correlated photons in Si nanowires was succesfully demonstrated in~\cite{Sharping2006}. Since then, several groups have reported Si-based entangled photon-pair sources based on various geometries that we present in the following.

\subsubsection{Straight nanowire waveguides}

\begin{figure}[htb]
\centerline{\includegraphics[width=0.5\textwidth]{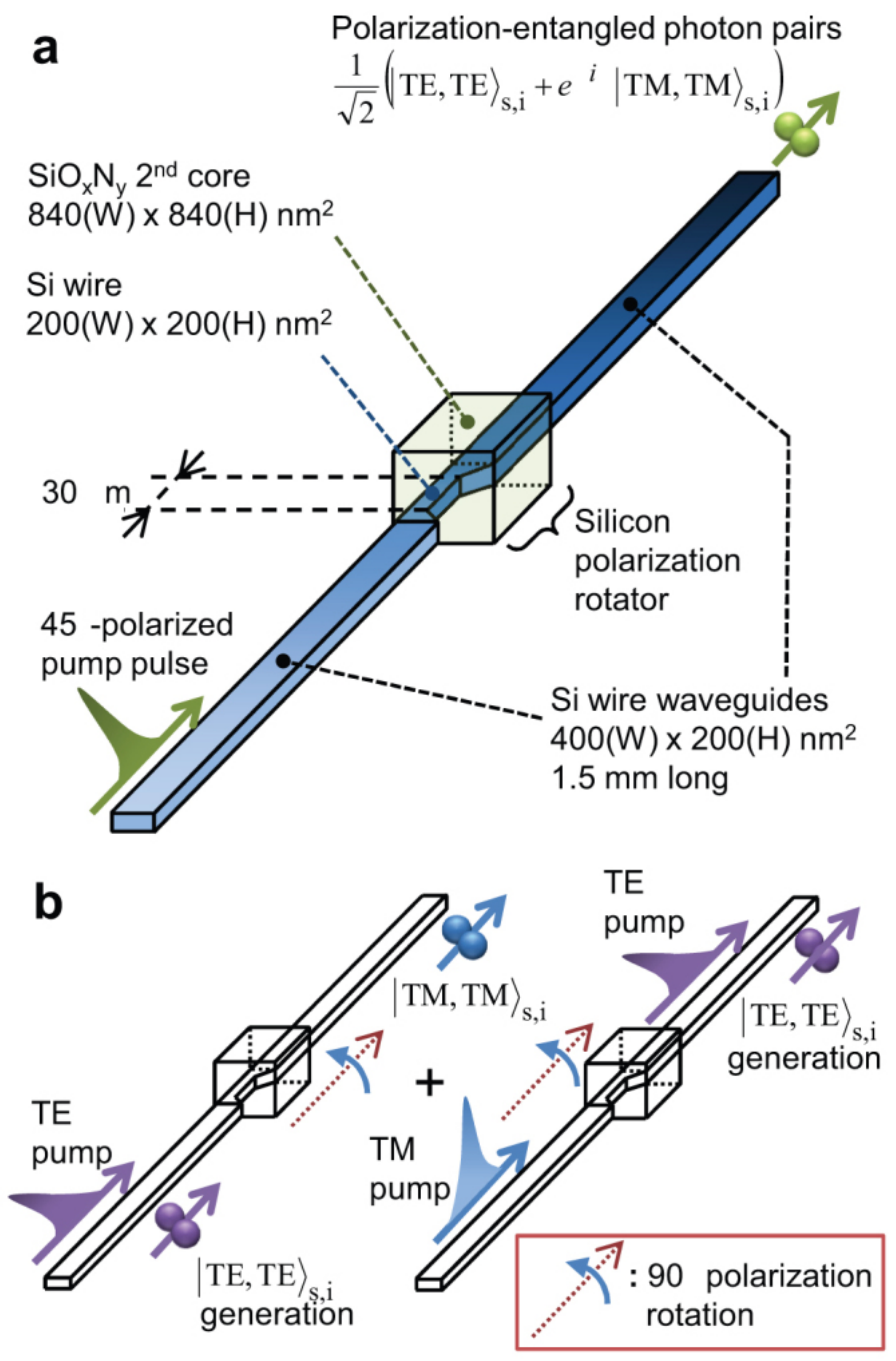}}
\caption{a) A monolithically integrated polarization-entanglement source consisting of a silicon-wire-based 90$^\circ$ polarization rotator sandwiched in between two nonlinear silicon nanowire waveguides. b) The device generates the polarization entanglement as a superposition of the two events: the TE component of the pump mode can generate $\ket{TE,TE}_{s,i}$ in the first nanowire waveguide which are then converted to $\ket{TM,TM}_{s,i}$ by the polarization rotator, or the TM component of the pump is converted to TE by the polarization rotator and can then generate $\ket{TM,TM}_{s,i}$ in the second nanowire waveguide. Nota bene: there is a typo in the original figure; one should read $30\,\mu m$ instead of $30\,m$ in panel a. (\copyright 2012 under Creative Commons, reproduced from~\cite{Matsuda2012})}
\label{SD11}
\end{figure}

The first entanglement generation with silicon wire waveguides has been reported in~\cite{Takesue2007}. The authors demonstrated time-bin entangled photons by pumping a centimetre-long waveguide with a CW telecom laser modulated into double pulses at a $100\,MHz$ repetition rate with an intensity modulator. A two-photon interference with a visibility larger than 0.73 was reported. This visibility has been increased up to 0.95 in~\cite{Harada2008} by employing mode size converters on both sides of the waveguide, thus reducing the outcoupling losses between the waveguide and external optical fibres. polarization entanglement has also been demonstrated by placing the silicon waveguide in a fibre loop~\cite{Takesue2008,Lee2008}. In this geometry the pump pulse was split into horizontal (H) and vertical (V) polarization components, which circulated in the loop in the counter-clockwise (CCW) and clockwise (CW) directions respectively, leading to the generation of a maximally entangled state $(\ket{H}_s\ket{H}_i +\ket{V}_s\ket{V}_i)/\sqrt{2}$. More recently, a fully integrated polarization entanglement source has been demonstrated~\cite{Matsuda2012} (Figure~\ref{SD11}). In this work, the authors used an integrated polarization rotator (a technology developed for telecommunication devices) in the midpoint of the nanowire in order to compensate the effect of the effective index difference between TE and TM modes and thus make the $\ket{TE,TE}_{s,i}$ and $\ket{TM,TM}_{s,i}$ photon pairs indistinguishable in the time degree of freedom. A full tomography of the generated state was performed, leading to a concurrence value $C = 0.88\pm0.02$.

\subsubsection{Microcavities}

\begin{figure}[htb]
\centerline{\includegraphics[width=1\textwidth]{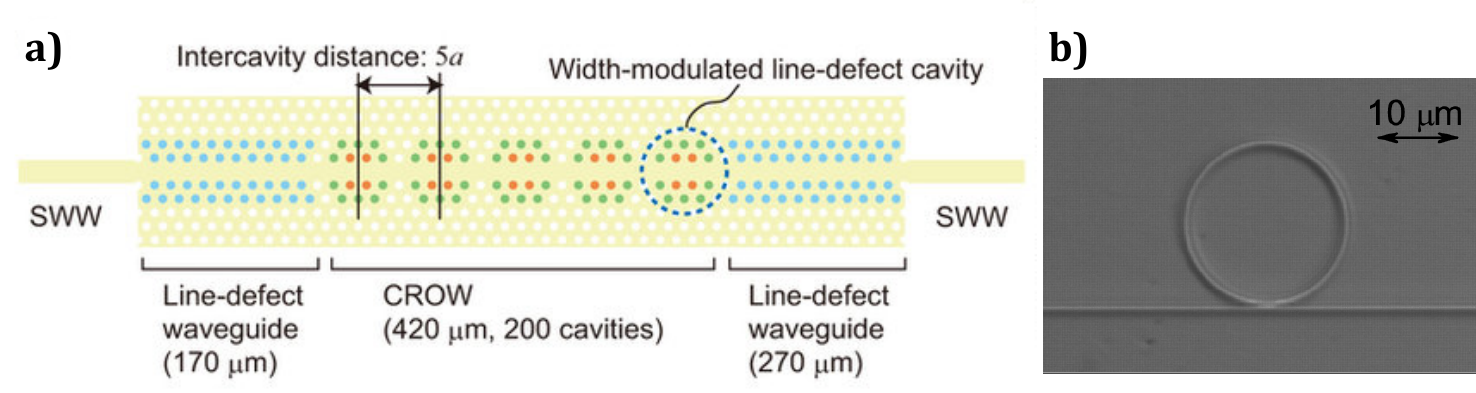}}
\caption{a) CROW structure used to generate time-bin entangled photon pairs (see the text for details). (\copyright 2014 under Creative Commons, reproduced from~\cite{Takesue2014}) b) Si ring resonator used to generated energy-time entangled photon pairs. (\copyright 2015 OSA, reproduced with permission from~\cite{Grassani2015})}
\label{SD12}
\end{figure}

In order to reduce the footprint of integrated sources from centimetre to micrometre scale, the technologies of coupled resonator optical waveguides (CROW) and ring resonators have been investigated.
A CROW is a one-dimensional array of identical optical cavities, where adjacent cavities are coupled to each other and form an extended mode along the waveguide. The transmission bandwidth of the resulting effective cavity is larger than the bandwidth of the individual cavities, while the group velocity is significantly reduced inside the transmission band. In~\cite{Takesue2014} a CROW based on a width-modulated line defect cavity in a silicon photonic crystal (PhC) was demonstrated (Figure~\ref{SD12}a), with a two-dimensional triangular lattice of air holes. The PhC section is integrated with silicon wire waveguides (SWW) allowing the optical addressing of the CROW. Thanks to SFWM enhanced by the slow-light effect in the device, the authors obtained an on-chip time-bin entangled photon-pair source with a device length of $420\,\mu m$. The limitation of the two-photon interference visibility in this work came from a high level of noise photons due to optical losses.
More recently, a silicon-on-silica ring resonator emitting energy-time entangled photons has been demonstrated~\cite{Grassani2015} (Figure~\ref{SD12}b), shrinking down the dimension of the device to an area of $300\,\mu m^2$. The authors performed a Franson-type experiment showing a violation of Bell’s inequality by more than seven standard deviations with an internal pair generation rate exceeding $10^7\,Hz$.

\subsubsection{Path-entanglement generation}

\begin{figure}[htb]
\centerline{\includegraphics[width=1\textwidth]{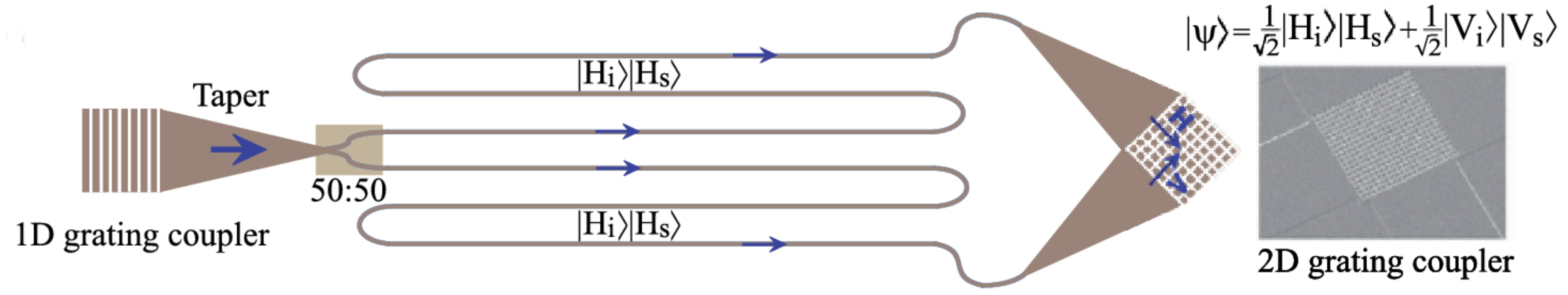}}
\caption{Scheme of a SOI chip producing polarization-entangled photon pairs (see the text for details). Inset (on the right): SEM image of the 2D grating coupler. (\copyright 2013 OSA, reproduced with permission from~\cite{Olislager2013}).}
\label{SD13}
\end{figure}

The technological maturity of silicon photonics in the telecom band has allowed increasing the complexity of quantum circuits by exploiting path entanglement.
Olislager and coworkers~\cite{Olislager2013} have demonstrated a chip producing polarization entanglement (Figure~\ref{SD13}). The key element of this source is a 2D coupler able to transform the path-entangled state generated in the chip into a polarization-entangled state at the output. In their scheme, a pump beam is coupled into the structure by using a 1D grating coupler followed by a taper. A 50/50 multimode coupler then splits the light into two silicon wire waveguides. Four-wave mixing in both waveguides leads to the generation of horizontally polarized photon pairs, and hence to the state $a'\ket{H}_{s1}\ket{H}_{i1} +b'\ket{H}_{s2}\ket{H}_{i2}$, where $s,i$ refer to signal and idler photons, and $1,2$ refer to the first and second waveguides. The 2D grating coupler then converts path entanglement into polarization entanglement, thus producing the state $a\ket{H}_s\ket{H}_i +b\ket{V}_s\ket{V}_i$ in the optical fibre at the output.

\begin{figure}[htb]
\centerline{\includegraphics[width=1\textwidth]{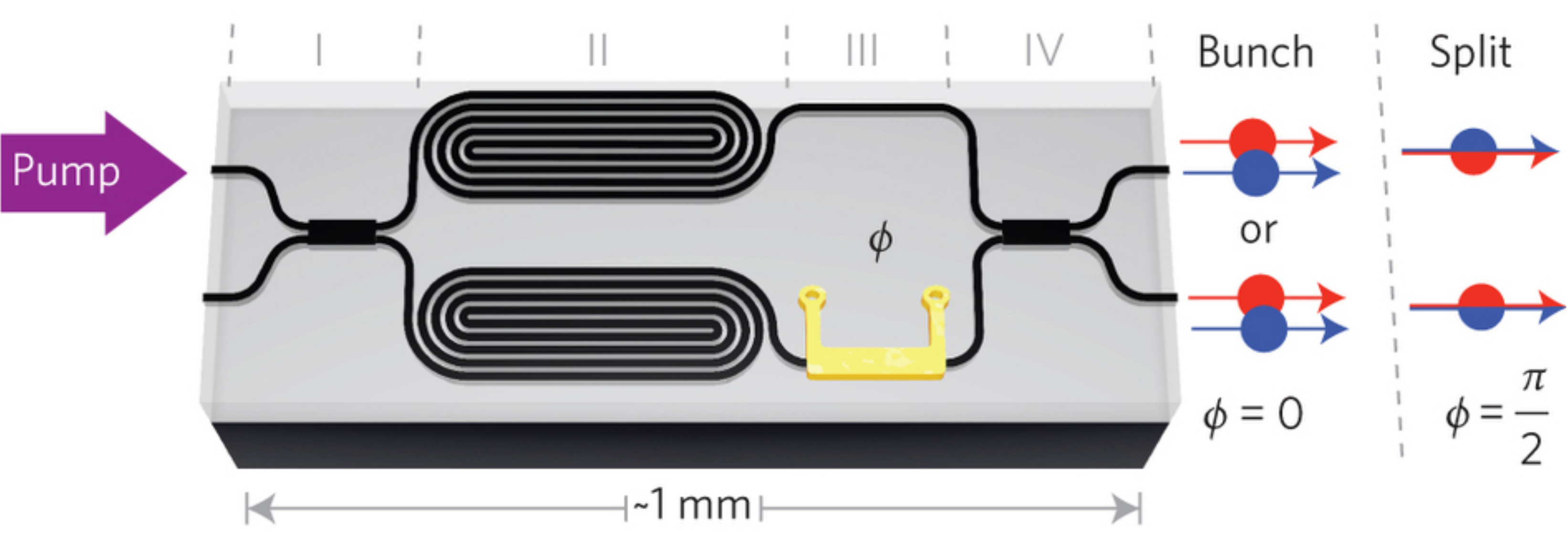}}
\caption{Scheme of a tunable SOI integrated source of path-entangled photon pairs. A bright pump laser is coupled to the SOI chip using a lensed optical fibre and on-chip spot-size converters (not shown). The pump is distributed between two modes via a multimode interference coupler (I), and excites the $\chi^{(3)}$ SFWM effect within each spiraled SOI waveguide source (II) to produce signal idler photon pairs in the two-photon entangled state $(\ket{20}-\ket{02})/\sqrt{2}$. The pairs are thermo-optically phase shifted ($\phi$, III) and interfered on a second coupler (IV) to yield either bunching or splitting over the two output modes, depending on $\phi$. (\copyright 2013 NPG, reproduced with permission from~\cite{Silverstone2013}).}
\label{SD14}
\end{figure}

In~\cite{Silverstone2013}, the authors demonstrate a chip combining two silicon FWM sources in an interferometer with a reconfigurable phase shifter (Figure~\ref{SD14}). The device is configured to create and manipulate both non-degenerate and degenerate path-entangled photon pairs. A two-photon interference visibility of $0.945\pm0.003$ (without background noise substraction) was observed on-chip.

\subsection{Quantum-dot-based sources}

In contrast to spontaneous parametric down-conversion sources, quantum dots offer the potential of generating single pairs of entangled photons on demand, that is, of generating one entangled photon pair per excitation pulse, and not more than one pair. This feature of quantum dots, together with the practical advantages of having a nanoscale source of entanglement, has stimulated the fabrication of various kinds of photonic microstructures with quantum dots, emitting entangled photons~\cite{BeveratosReview}.

The Stranski-Krastanow growth mode is a common method of fabricating quantum dots for quantum optics applications. A thin wetting layer of InGaAs (typically) is deposited by molecular-beam epitaxy (MBE) or metal-organic vapor phase epitaxy (MOVPE) on a GaAs surface. Because of the crystal lattice mismatch with the GaAs, strain builds up as the InGaAs layer becomes thicker. When a thickness of about 1.5 monolayers has been reached, it becomes energetically more favorable for the InGaAs to form islands~\cite{Masumoto2002}. These islands are called self-assembled quantum dots and they appear at random positions on the GaAs substrate. A GaAs capping layer is subsequently deposited to protect the quantum dots from oxidation. The capping layer is also necessary to create the electrical potential for quantum confinement of electrons and holes in the quantum dot. Typically, self-assembled quantum dots have dimensions of $5-10\,nm$ in height and a few tens of nanometres in diameter. For typical InGaAs self-assembled quantum dots, the emitted photons have wavelengths around $900\,nm$.

\subsubsection{Entanglement generation via the biexciton-exciton cascade}

The cascaded photon emission from a biexciton state, introduced in Section~2, can be used to generate entangled photon pairs from a quantum dot. Here, we will address four schemes allowing to do so, and their prerequisites.

\vspace{0.2cm}
\noindent \textbullet~\textit{polarization-entangled photon pairs from a quantum dot with zero fine-structure splitting.}

\begin{figure}[htb]
\centerline{\includegraphics[width=0.7\textwidth]{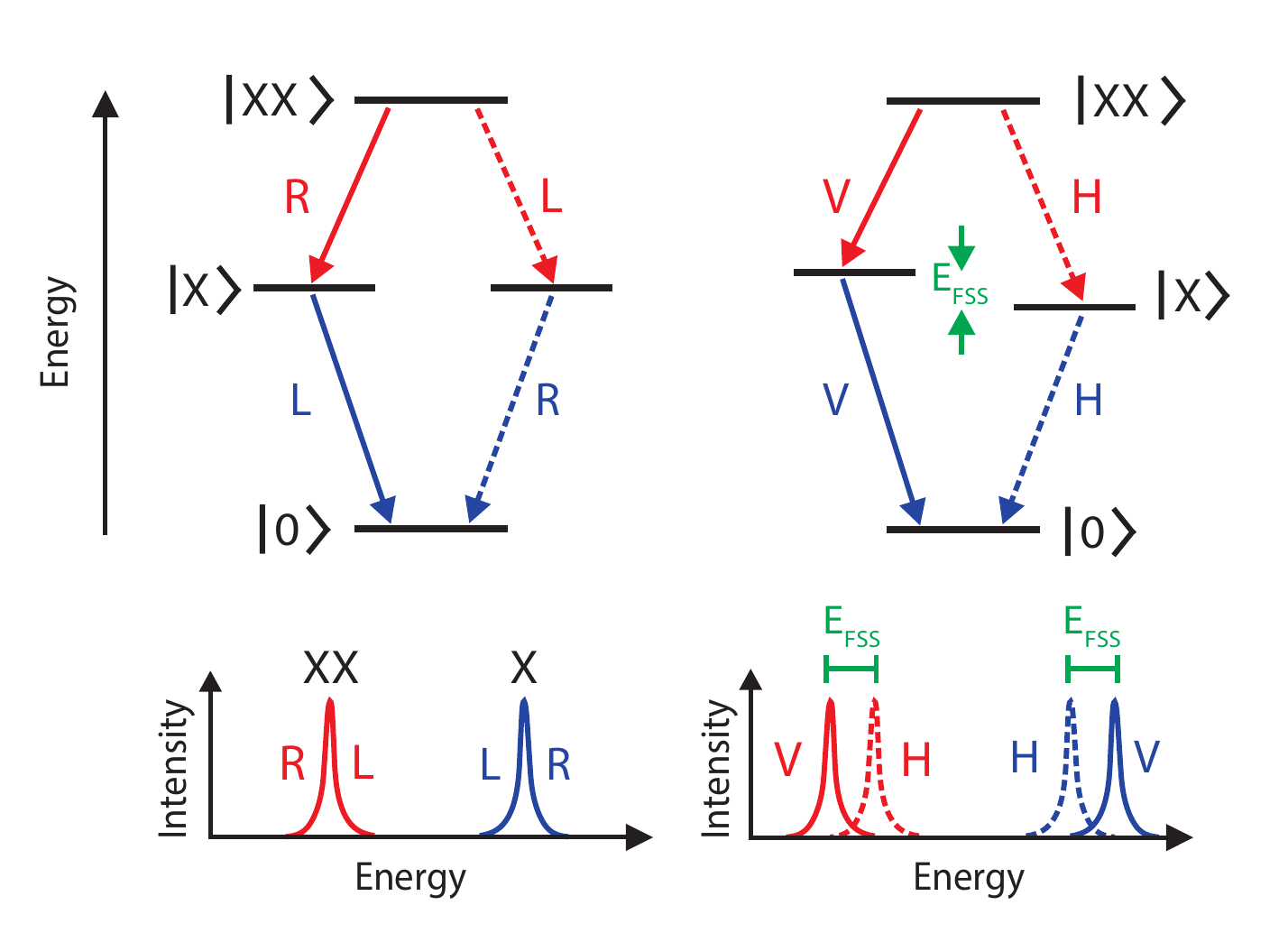}}
\caption{Biexciton-exciton cascade for a quantum dot with zero fine-structure splitting (left) and non-zero fine-structure splitting (right). The top row depicts the energy level schemes and below the corresponding emission spectra are depicted.}
\label{KDJ2}
\end{figure}

The most common scheme to generate polarization-entangled photon pairs from the biexciton-exciton cascade of a quantum dot was presented by Benson \textit{et al.}~\cite{Benson2000} in the year 2000. The scheme takes advantage of the Pauli exclusion in the s-shell of the quantum dot, resulting in a zero-spin bound biexciton state when the s-shell is fully occupied. The decay to the ground state from the biexciton over the exciton state takes place by the emission of two cascaded photons, with zero total angular momentum. Due to the optical selection rules, the recombination of one electron-hole pair from the biexciton state $\ket{XX}$ results in the emission of a left (L) or right (R) circularly polarized biexciton photon (XX), depending on the spin configuration of the recombining electron-hole pair. The degenerate exciton state $\ket{X}$ is in a superposition state and emits an exciton photon (X) with opposite circular polarization (R or L) with respect to the polarization of the previously emitted XX photon. The polarization state of the photon pair is a maximally entangled Bell state, since the wave function of the polarization state of the photon pair cannot be separated into a product state of the wave functions of each polarization state of the individual photons. The entangled polarization state can be written as:
\begin{equation}
\fl
\ket{\phi^+}=\frac{1}{\sqrt{2}}\big(\ket{R}_{XX}\ket{L}_{X}+\ket{L}_{XX}\ket{R}_{X}\big) = \frac{1}{\sqrt{2}}\big(\ket{H}_{XX}\ket{H}_{X}+\ket{V}_{XX}\ket{V}_{X}\big).
\end{equation}
The state has the same form in any basis and can be rewritten e.g. in the rectilinear basis, where H (V) refers to the horizontal (vertical) polarization.

In 2006 the first experimental demonstrations of the emission of entangled photon pairs from a quantum dot were performed~\cite{Akopian2006,Young2006}. Since then, a lot of effort has been made to improve the quantum dot-based entangled photon pair sources~\cite{Hafenbrak2007, Dousse2010,Juska2013,Kuroda2013}. Recently, the on-demand generation of polarization-entangled photon pairs~\cite{Muller2014} was achieved, using a two-photon excitation scheme~\cite{Brunner1994,Stufler2006} to resonantly excite the biexciton state.

The cascaded process is schematically depicted on the left part of Figure~\ref{KDJ2}: the energy level scheme and the corresponding photoluminescence spectrum of the quantum dot are shown. The L and R exciton (biexciton) photons have the same energy, since both spin configurations of the exciton are energetically degenerate. However, most commonly, the four exciton states are nondegenerate, due to the coupling between different electron and hole spin configurations~\cite{Bester2003}. The two previously degenerate bright exciton states couple, forming new eigenstates with different energies (right part of of Figure~\ref{KDJ2}). The splitting between the bright exciton levels is called the fine-structure splitting (FSS)~\cite{Gammon1996}. To understand the origin of this splitting we can describe the electron hole exchange interaction by the following Hamiltonian $\hat{H}_{exch}$~\cite{Bayer2002}:
\begin{equation}
\fl
\hat{H}_{exch} = - \sum_{i=x,y,z} (a_i \hat{S}_{h,i}\cdot\hat{S}_{e,i} + b_i \hat{S}_{h,i}^3\cdot\hat{S}_{e,i}),
\end{equation}
where $\hat{S}_{h,i}$ ($\hat{S}_{e,i}$) stand for the hole (electron) spin operator and $a_i$ and $b_i$ are the spin-spin coupling constants in the three spatial axes $i=x,y,z$. Their magnitude depends on the confining potential in the specific spatial axis. For example a reduction of the in-plane rotational symmetry of the confining potential ($<\text{D}_{\text{2D}}$ symmetry) results in different spin-spin coupling constants $b_x\neq b_y$. This difference leads to the splitting of the bright $\ket{X}$ states. Note that, for a qualitative non-atomistic description of the splitting, long-range exchange interactions have to be also taken into account, leading to a bright exciton fine-structure splitting of:
\begin{equation}
\fl
E_{FSS} = \frac{3}{8}(b_x-b_y) + (\gamma_x - \gamma_y),
\end{equation}
where $\gamma_x$ and $\gamma_y$ denote the contributions of the long-range exchange interaction. A complete analysis is given in Ref.~\cite{Bayer2002}. In the case of the $\ket{XX}$ state, both electron and hole spins are in a singlet state, resulting in no fine structure for the $\ket{XX}$ state. However, one will still observe the fine-structure splitting in the emitted biexciton photons' energy since the $\ket{XX}$ state recombines into the intermediate $\ket{X}$ state. Therefore, the excitonic FSS can also be measured in the spectrum of the biexciton photons, as shown in the schematic spectrum of Figure~\ref{KDJ2} on the bottom right.
The fine-structure splitting of the $\ket{X}$ state leads to an exciton-spin precession, which directly affects the polarization state:
\begin{equation}
\label{EqKlaus}
\fl
\ket{\psi} = \frac{1}{\sqrt{2}}(\ket{H}_{XX}\ket{H}_{X}+e^{i\tau_xE_{FSS}/\hbar}\ket{V}_{XX}\ket{V}_{X}),
\end{equation}
where $\tau_x$ is the time interval between the emission of the two photons. This time-evolving state depends on the product of this time interval and the fine-structure splitting. A comprehensive study on the time-evolving entangled state can be found in the work of Ward \textit{et al.}~\cite{Ward2014}, which introduces a time-dependent formalism for entanglement measurements. Such a time-evolving entanglement cannot be observed in conventional time-integrated measurements since, over time, the instantaneous superpositions cancel out with those with opposite phase~\cite{Stevenson2008}. This makes quantum dots with fine-structure splitting typically unpractical for entangled photon pairs generation. However, by employing quantum feedback, it is possible to increase the range of acceptable fine-structure splitting and still generate a substantially strong entanglement~\cite{Hein2014}.
In addition, there are several possibilities to reduce the fine-structure splitting, either by different growth methods or by post-growth tuning techniques. For example, one can fabricate quantum dots with an intrinsic symmetric confining potential~\cite{Juska2013,Kuroda2013,Huo2013} or use thermal annealing~\cite{Langbein2004} to restore the rotational in-plane symmetry of the confining potential. Successful post-growth tuning techniques are: electric field-~\cite{Kowalik2005,Gerardot2007,Marcet2010}, magnetic field-~\cite{Stevenson2006}, and strain tuning~\cite{Seidl2006,Rastelli2012}, or a combination of those~\cite{Trotta2012b}.

\vspace{0.2cm}
\noindent \textbullet~\textit{polarization-entangled photon pairs from a quantum dot via time reordering.}

\begin{figure}[htb]
\centerline{\includegraphics[width=0.7\textwidth]{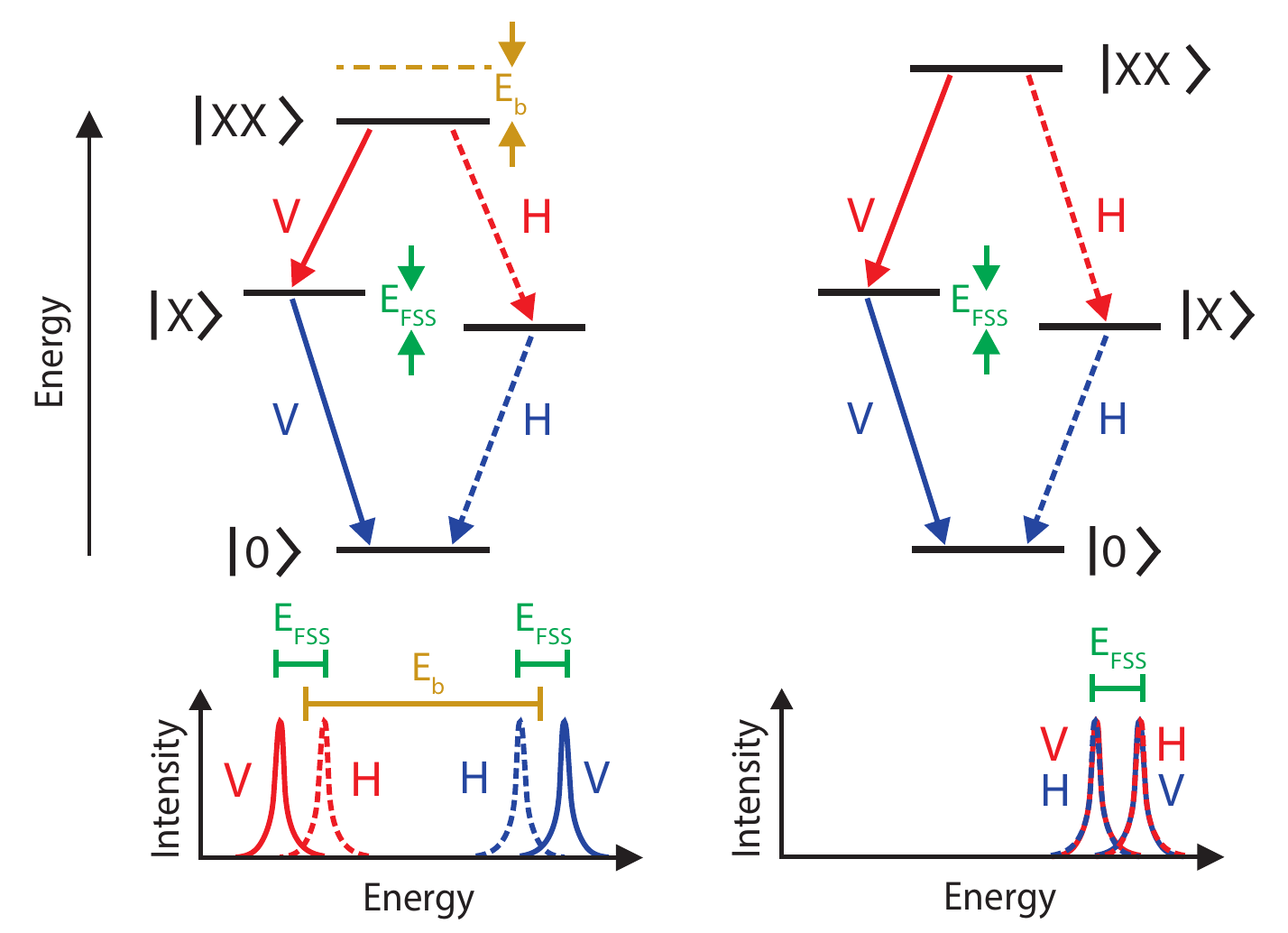}}
\caption{Biexciton-exciton cascade for a quantum dot with biexciton binding energy $E_b$ (left) and $E_b=0$ (right). Top row depicts the energy level schemes and below are the corresponding emission spectra. In case of $E_b=0$ the H (V) polarized XX photon energetically overlaps with the V (H) polarized X photon.}
\label{KDJ3}
\end{figure}

Instead of using hard-to-find quantum dots with zero fine-structure splitting for the generation of polarization-entangled photon pairs, one can use a quantum dot with zero biexciton binding energy $E_b$ instead. This scheme was initially proposed in~\cite{Reimer2007} and a full theoretical description of the scheme and of its feasibility followed shortly after~\cite{Avron2008}. Figure~\ref{KDJ3} compares the required level scheme of such a quantum dot with the level scheme a normal quantum dot with FSS. In the corresponding spectrum (on the bottom of figure~\ref{KDJ3}), the effect of $E_b=0$ is visible: the H (V) biexciton photon has now the same energy as the V (H) exciton photon, leading to a coincidence of colours across the recombination pathways, rather than within the recombination pathways. 
However, one can still distinguish these photons due to their different arrival time, since the biexciton photon is always emitted before the exciton photon. This information has to be erased to generate polarization entanglement between the biexciton and exciton photon. 
In addition, to achieve a large degree of entanglement, not only the colours have to match perfectly: the generation also depends on the life time $\tau$ of the $\ket{XX}$ and $\ket{X}$ states. Ideally one would need $\tau_{X}/\tau_{XX}=0$, however for quantum dots $\tau_{X}/\tau_{XX}\approx 2$ typically holds. Still, substantial entanglement can be created via the biexciton-exciton cascade~\cite{Avron2008}. The effect of the different life times of the involved states on the entanglement generation was in-depth theoretically analysed and a maximum concurrence $C = 0.736$ was found when both biexciton recombination channels as well as both exciton recombination channels have the same decay time and $\tau_{X} << \tau_{XX}$~\cite{Troiani2008}. With additional spectral filtering, at the cost of efficiency, significantly larger values of entanglement with the time reordering scheme can be achieved~\cite{Pathak2009}.

The described scheme relies therefore on two important steps: i) tuning of the biexciton binding energy $E_b$ to zero and ii) time reordering of the emitted photons. Tuning $E_b$ to zero can be realized, for example, by the application of an electric field~\cite{Reimer2011} or of a biaxial strain perpendicular to the quantum dot growth axis~\cite{Ding2010,Trotta2012a}, or a combination of both, which would additionally allow to tune the emission energy of the emitted entangled photons~\cite{Trotta2013}. However, efficient time reordering of the biexciton and exciton photons is experimentally challenging. One initial idea was to first go through a linear dispersive grating to separate the biexciton and exciton photons. Then use two polarizing beam splitters and different delay lines to reorder the photons and recombine them afterwards. And then, once more, go through the grating to avoid introducing a chirp~\cite{Avron2008}. Another recent idea is to slow down light in an atomic vapor. The accumulated delay through the atomic vapor depends on the photon energy and could be tailored to successfully time reorder the biexciton and exciton photons~\cite{Wildmann2015}.
To the best of our knowledge, an experimental realization of the complete scheme, including the tuning of $E_b$ to zero and the time reordering of the emitted photons, has not been demonstrated yet.

\vspace{0.2cm}
\noindent \textbullet~\textit{Time-bin entangled photon pairs from a quantum dot.}

\begin{figure}[htb]
\centerline{\includegraphics[width=0.7\textwidth]{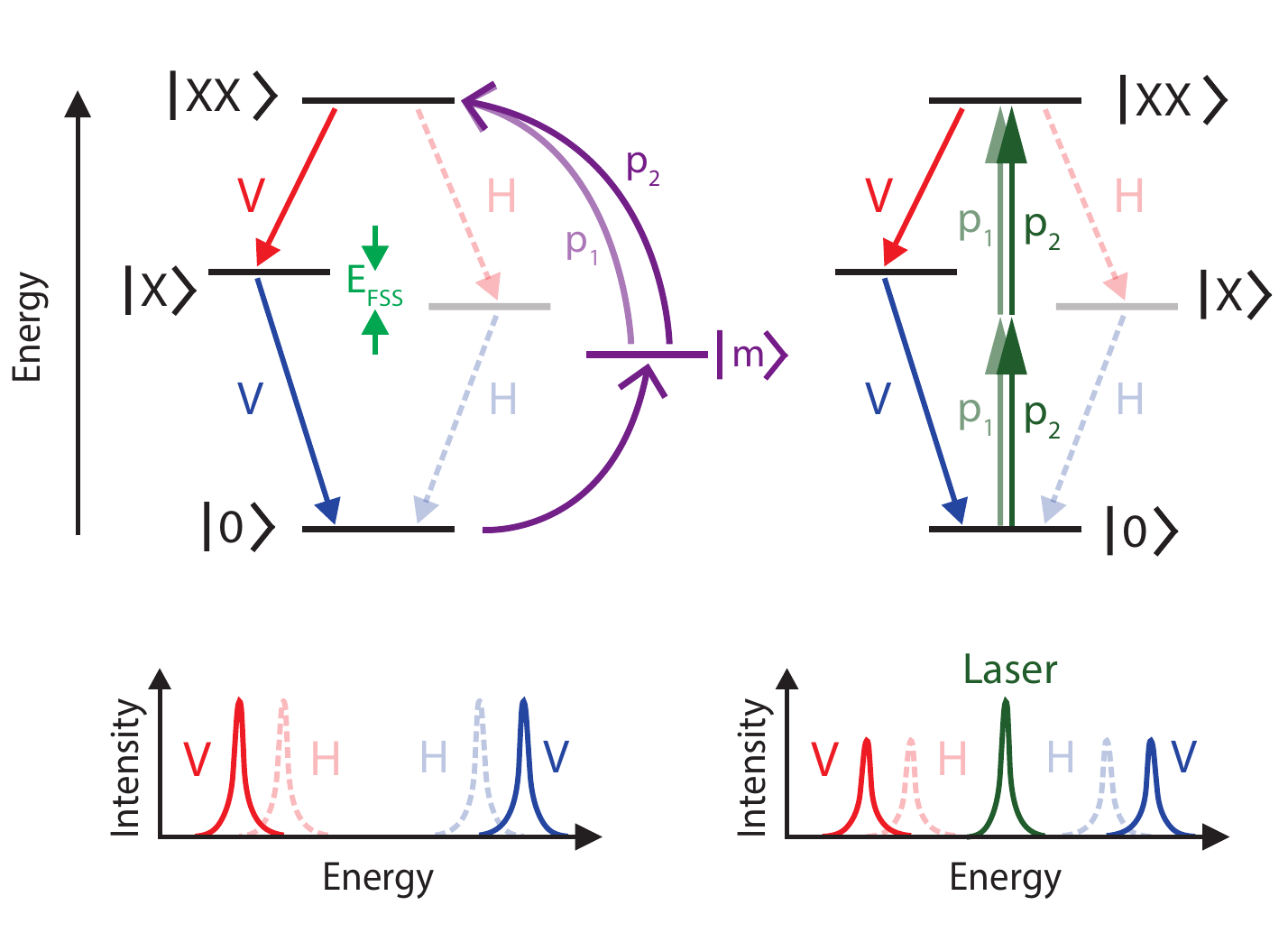}}
\caption{Schematic excitation schemes to generate time-bin entangled photon pairs from the biexciton-exciton cascade. Either a long-lived metastable state $\ket{m}$ (left) or a resonant two-photon excitation scheme (right) is used for the subsequent excitation ($p_1$ and $p_2$) of the $\ket{XX}$ state.}
\label{KDJ4}
\end{figure}

The initial scheme for generating time-bin entangled photons from quantum dots was proposed by C. Simon and J.-P. Poizat in 2005~\cite{Simon2005}. It has no stringent requirements for the fine-structure splitting. The basic idea is to excite the quantum dot in the $\ket{XX}$ state with a probability $p_1=0.5$ with a first pulse and, then, apply another excitation pulse which ideally populates the $\ket{XX}$ state with a probability $p_2=1$. This results in either a biexciton-exciton photon pair emitted after the early excitation or after the late excitation. The idealized entangled two-photon state can be written as:
\begin{equation}
\fl
\ket{\psi} = \sqrt{p_1}\ket{\textrm{early}}_{XX}\ket{\textrm{early}}_{X}+e^{i\phi}\sqrt{(1-p_1)p_2}\ket{\textrm{late}}_{XX}\ket{\textrm{late}}_{X},
\end{equation}
where $\phi$ denotes the relative phase between the early and the late excitation pulses. From the equation it is clear that the correct ratio of $p_1$ and $p_2$ is crucial to generate a maximally entangled state. Ideally, one wishes to avoid excitation of the quantum dot by both excitation pulses. This requires the preparation of the quantum dot into a long lived or metastable state $\ket{m}$, e.g. a dark exciton~\cite{Poem2010} or an off-resonant bright exciton in a photonic band-gap structure~\cite{Hennessy2007}.

The working principle is depicted in Figure~\ref{KDJ4}, on the left. Starting from the ground state, the system has to be excited in the metastable state $\ket{m}$, which is non-trivial. Afterwards the $\ket{XX}$ state is prepared with two consecutive excitation pulses. The scheme only uses one recombination pathway of the biexciton-exciton cascade. The other (shaded) is typically discarded via polarization filtering to have the photon pair in a well-defined polarization mode. Ideally, a cavity which suppresses the other recombination pathway should be employed to maintain a high generation efficiency.
Another approach is to use a coherent excitation scheme from the metastable state, using cavity-assisted piecewise adiabatic passage~\cite{Pathak2011}. This would provide an efficient way for the population transfer to the $\ket{XX}$ state. Giving small values of pure dephasings, the cavity-assisted scheme might result in the generation of on-demand time-bin entangled photon pairs.

Both described approaches require the addressability of a metastable or long-lived state. This is experimentally still a challenge. Instead, the group of G. Weihs generated time-bin entangled photon pairs from the biexciton-exciton cascade~\cite{Jayakumar2014} through resonant excitation of the biexcition via a two-photon excitation scheme, depicted on the right of Figure~\ref{KDJ4}. The laser is detuned from the $\ket{X}$ state and only excites the quantum dot resonantly in the $\ket{XX}$ state, via a two-photon process (two green arrows) with a virtual state. By using a two-pulse sequence and by keeping the excitation probability small for the early excitation pulse, this allowed them to generate, for the first time, time-bin entangled photons from a quantum dot. Their approach can generate a maximally entangled state but does not suppress the multi-photon emission caused by double excitation. Therefore, they kept the excitation probability fairly low. Thus on-demand generation with this scheme would be troublesome.
In contrast, the multi-photon emission is strongly suppressed in an alternative approach, where, first, single pairs of polarization-entangled photons are created from a quantum dot with a small fine-structure splitting, and where, subsequently, the polarization entanglement is converted into time-bin entanglement~\cite{Versteegh2015}. This approach allows for on-demand generation of entangled pairs.

\vspace{0.2cm}
\noindent \textbullet~\textit{polarization-entangled photon pairs via two-photon emission from the biexciton.}
\vspace{0.2cm}

Similarly to the above mentioned two-photon excitation of the biexciton state via a virtual level, one can reverse the process to simultaneously generate two photons. The $\ket{XX}$ state recombines with the help of a virtual state directly to the quantum dot ground state, emitting a polarization-entangled photon pair~\cite{Schumacher2012}. Such schemes where the intermediate $\ket{X}$ state is jumped over are also called leapfrog processes~\cite{Gonzalez-Tudela2013}. Typically, this two-photon emission process is weak compared to the radiative biexciton-exciton cascade. However, with the help of cavity quantum electrodynamics, one can tune the virtual state into resonance with a strong cavity mode~\cite{delValle2013}. This increases the probability of a two-photon process compared to the normal cascaded emission. In 2011, Ota \textit{et al.}~\cite{Ota2011} demonstrated such a two-photon spontaneous emission of a single quantum dot by embedding the quantum dot in a photonic crystal nanocavity.

In this entanglement scheme the $\ket{XX}$ state has two competing recombination pathways, either via the biexciton-exciton cascade or via the spontaneous two-photon emission generating entangled photon pairs. The probability to generate polarization-entangled photon pairs depends on the coupling strength to the cavity mode. A narrower cavity resonance reduces the coupling to the detuned biexciton-exciton cascade and enhances the direct two-photon transition through the Purcell effect~\cite{Purcell1946}.  
However, the cavity mode can be seen as a filter, distilling the entanglement at the expense of the brightness. An on-demand generation of entangled photon pairs would only be possible if one could ensure that every biexciton excitation recombines over the two-photon process. Up to now, no experimental demonstration of entangled photon pairs from such an entanglement scheme has been reported.

\subsubsection{Quantum dots in photonic microstructures} 

For practical purposes in quantum communications, it is desirable to generate entangled photons with a high rate. One problem of self-assembled quantum dots is that, because of the high refractive index of the semiconductor material ($n\approx 3.6$ for GaAs around $900\,nm$), only a very small portion (a few percents) of the light emitted from the quantum dot can escape the sample into the air and be captured by a high-NA lens positioned very close to the sample. 
A solution to overcome this problem is to grow distributed Bragg reflectors on the bottom and on the top of the structure, thus creating a semiconductor planar microcavity. The first demonstrations of entangled-photon-pair generation with quantum dots via the biexciton-exciton cascade were all done with self-assembled quantum dots in such microcavity structures~\cite{Young2006,Akopian2006,Hafenbrak2007}. When the resonance of the microcavity overlaps with the emission frequency of the quantum dot, the light extraction efficiency can be enhanced in three ways. i) The microcavity makes the emission more directional. ii) Usually, the number of distributed Bragg reflectors is chosen to be smaller on the top side than on the back side: this enhances the light emission from the top side. iii) The microcavity can increase the density of available photon modes into which the excited quantum dot can emit a photon, thereby the spontaneous emission rate increases according to Fermi's golden rule: this effect is called the Purcell effect~\cite{Purcell1946}. The Purcell effect occurs in microcavities with a high quality factor, small optical mode volumes, and good spatial and spectral overlaps between the cavity modes and the quantum dots. When the spontaneous emission rate is increased in this manner, the nonradiative decay and the effects on the photon emission of other processes, such as charge capture, spin flips and dephasing, are suppressed. Using quantum dots embedded in pillar microcavities, G\'{e}rard \textit{et al.}~\cite{Gerard1998} obtained favorable conditions for the Purcell effect and increased the spontaneous emission rate by a factor 5. Santori \textit{et al.}~\cite{Santori2002} produced indistinguishable single photons in such pillar microcavities. However, in the case of entanglement generation via the biexciton-exciton cascade, Purcell enhancement is a limited resource. Indeed, the biexciton photon and the exciton photon, in general, have different frequencies so only moderate quality factors can be chosen in order to keep both frequencies within the resonance of the cavity~\cite{Rousset2016}.

\begin{figure}[htb]
\centerline{\includegraphics[width=0.2\textwidth]{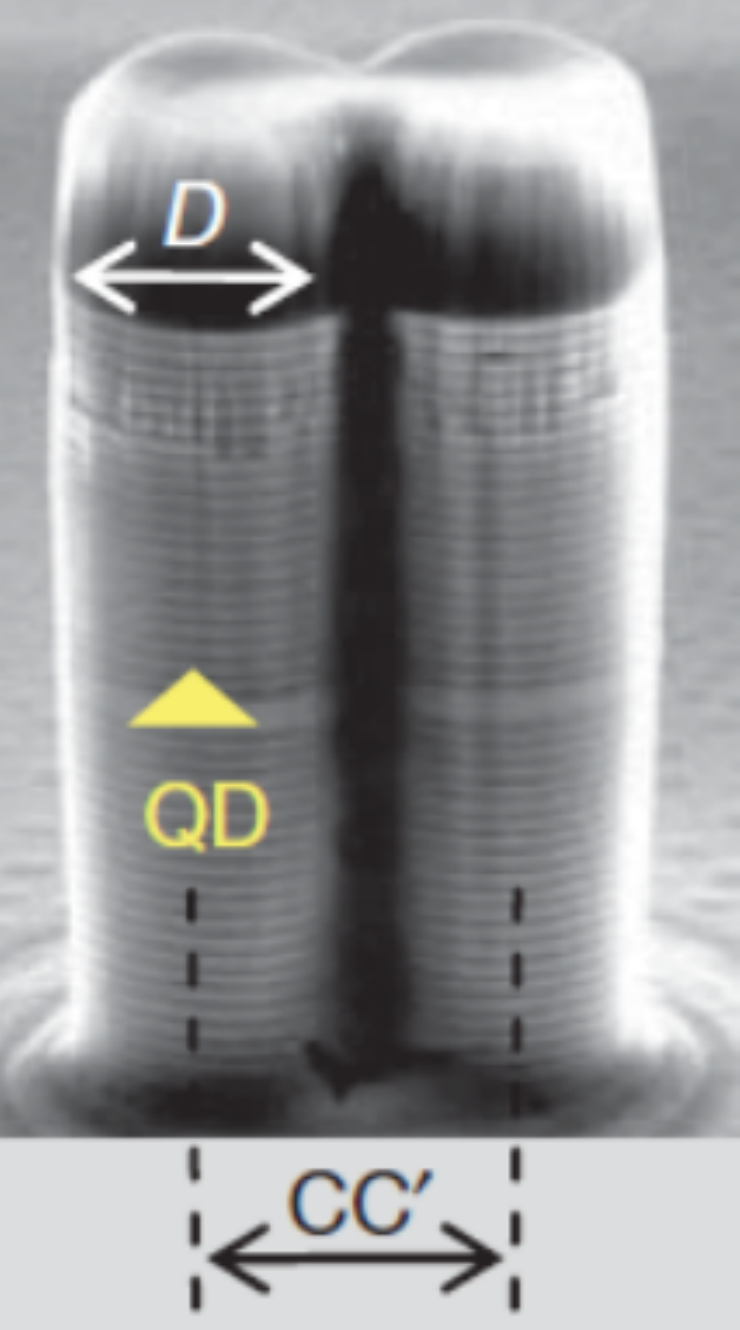}}
\caption{Double-micropillar structure, also called ‘photonic molecule’. Each pillar contains a microcavity and a quantum dot (QD) is present in one of the pillars. The diameter of the pillars $D$ and the distance between the pillar centers $CC’$ are chosen in such a way that the XX emission and the X emission are each exactly on resonance with subsequent modes of the photonic molecule. (\copyright 2010 NPG, reproduced with permission from~\cite{Dousse2010})}
\label{MV2}
\end{figure}

A further substantial improvement of the extraction efficiency of entangled photons was obtained by Dousse \textit{et al.} by etching a double-micropillar structure out of a semiconductor planar microcavity, the quantum dot being in one of the two pillars~\cite{Dousse2010} (Figure~\ref{MV2}). The diameters of the pillars and the center-to-center distance between the pillars can be tuned in such a way that the biexciton emission is in resonance with a cavity mode in one of the pillars, while the exciton emission is in resonance with a cavity mode in the other pillar. In this way, a Purcell factor of 3-5 was achieved and a collection efficiency into the first lens of 0.12 was obtained for each pair, a large improvement compared to the first demonstrations of entangled photons from quantum dots. In addition, thanks to the Purcell effect, the X lifetime was shortened and the homogeneous linewidth of the X transition was increased beyond the FSS, erasing the 'which path' information encoded in the energy of the emitted photons. The measured concurrence was 0.267 without temporal selection, which demonstrates entanglement, and 0.387 with temporal selection. The measured fidelity to a maximally entangled state was 0.63 without temporal selection.

Improving the extraction efficiency is easier in the case of path entanglement generation via HOM interference (see Section~2.4), since all photons have the same optical frequency in this case, so one could work with a single cavity with a high quality factor. By fabricating a microlens on top of a quantum dot, Gschrey \textit{et al.}~\cite{Gschrey2015} improved the extraction efficiency, while maintaining a strong indistinguishability of the photons. Very high brightness and indistinguishability have been obtained with quantum dots in pillar microcavities~\cite{Gazzano2013}, especially with resonant excitation~\cite{Somaschi2016,He2016,Wang2016,Ding2016,Loredo2016b}. A comparison in terms of brightness and indistinguishability of several quantum dot and parametric down-conversion sources is given in~\cite{Somaschi2016}.

\begin{figure}[htb]
\centerline{\includegraphics[width=0.4\textwidth]{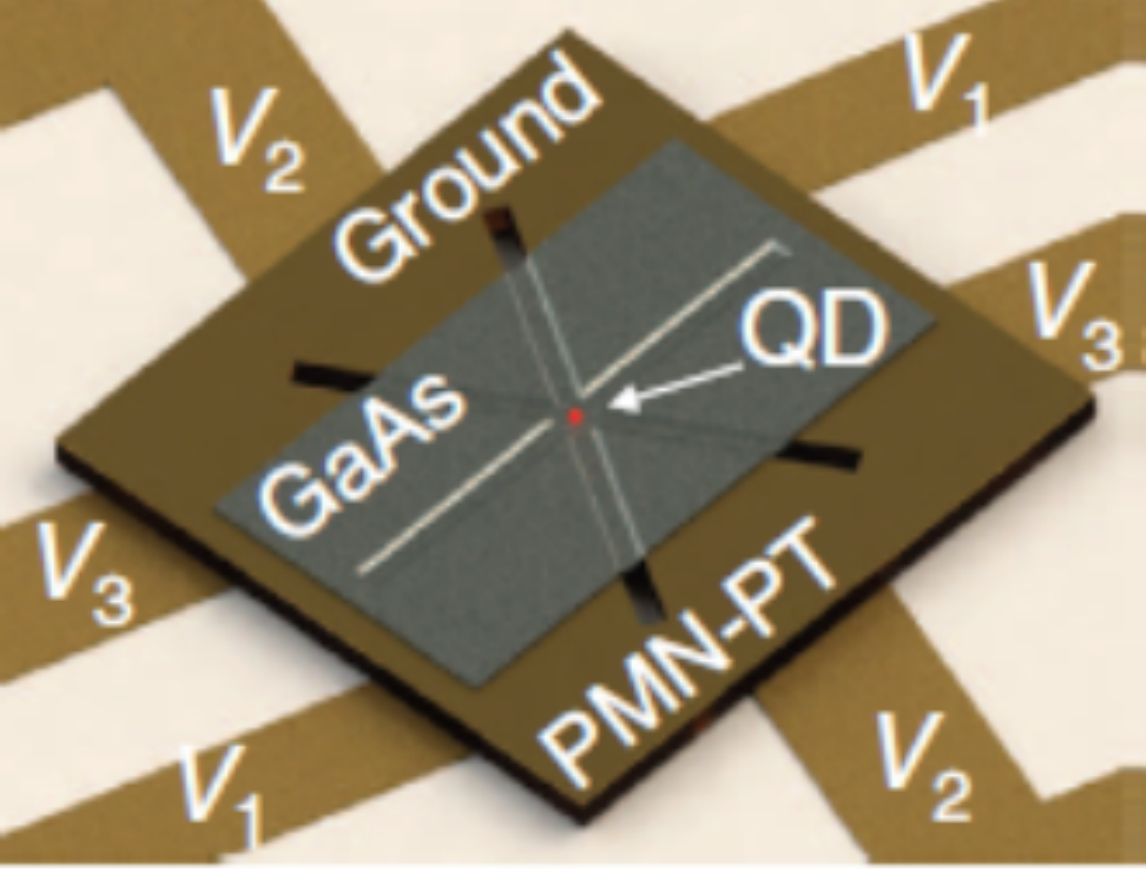}}
\caption{Quantum dot source of entangled photons with zero fine-structure splitting. By tuning the voltage on the six legs of this device, the strain of the nanomembrane (the grey part) is controlled. This way, the fine-structure splitting of a quantum dot inside this nanomembrane can be tuned to zero. (\copyright 2016 under Creative Commons, reproduced from~\cite{Trotta2016})}
\label{MV3}
\end{figure}

For the rest of this section we only discuss quantum dots in photonic microstructures for entanglement generation via the biexciton-exciton cascade. So far, all discussed sources of entangled photons were optically driven by laser excitation. For practical applications, such as a linear optical quantum computing, it is important to have an electrically driven entanglement source. Such a source, an entangled-light-emitting diode, was built by Salter \textit{et al.}~\cite{Salter2010}. It consisted of a layer of InAs quantum dots, embedded in a p-i-n doped planar microcavity, which emitted entangled photons from the biexciton-exciton cascade. The indistinguishability of photons from two subsequent excitation pulses was also demonstrated~\cite{Stevenson2012}. A maximum time-gated fidelity to a maximally entangled state of $0.87 \pm 0.04$ was reported.\\
As discussed in the previous section, a major limitation of the biexciton-exciton entanglement scheme is the fine-structure splitting between the two exciton states in the quantum dot, which occurs in quantum dots where the quantum confinement of the electrons and holes is not symmetric. Because of the resulting precession of the exciton spin, the measured entanglement is reduced. Entanglement may be recovered by spectral selection, selecting only the overlapping parts of the emission lines~\cite{Akopian2006}, or by temporal selection, i.e. selecting only photon pairs where the exciton decayed very quickly after the biexciton, so that the exciton spin precession remained small~\cite{Salter2010}. For the latter approach, one needs to select a temporal window smaller than $\hbar/E_{FSS}$ (see Equation~\ref{EqKlaus}). Both methods obviously lead to large losses as the majority of the emitted photon pairs is rejected. Another solution is to find a quantum dot which, by chance, has been formed symmetrically, and therefore exhibits no fine-structure splitting~\cite{Hafenbrak2007}.\\
Trotta \textit{et al.}~\cite{Trotta2014,Trotta2016} demonstrated a device, with quantum dots inside the intrinsic region of a p-i-n structure, where piezoelectric strain tuning was used to remove the fine-structure splitting, so that almost any quantum dot in the device could be used for the generation of entangled photons (Figure~\ref{MV3}). Similar devices were demonstrated by~\cite{Zhang2015,Chen2016}. Other successful strategies to remove the fine-structure splitting involve applying a magnetic field~\cite{Stevenson2006}, or a continuous wave laser field and making use of the optical Stark effect~\cite{Muller2009}, or applying a static electric field and making use of the quantum-confined Stark effect~\cite{Bennett2010}. By adding a strain relaxing layer to the structure, and using the quantum-confined Stark effect, Ward \textit{et al.}~\cite{Ward2014} were able to extend the wavelength of the emitted entangled photons to a telecommunication band (around $1300\,nm$). Details on the various post-growth methods to reduce the fine-structure splitting can be found in the review paper by Plumhof \textit{et al.}~\cite{Plumhof2012}.

\begin{figure}[htb]
\centerline{\includegraphics[width=0.6\textwidth]{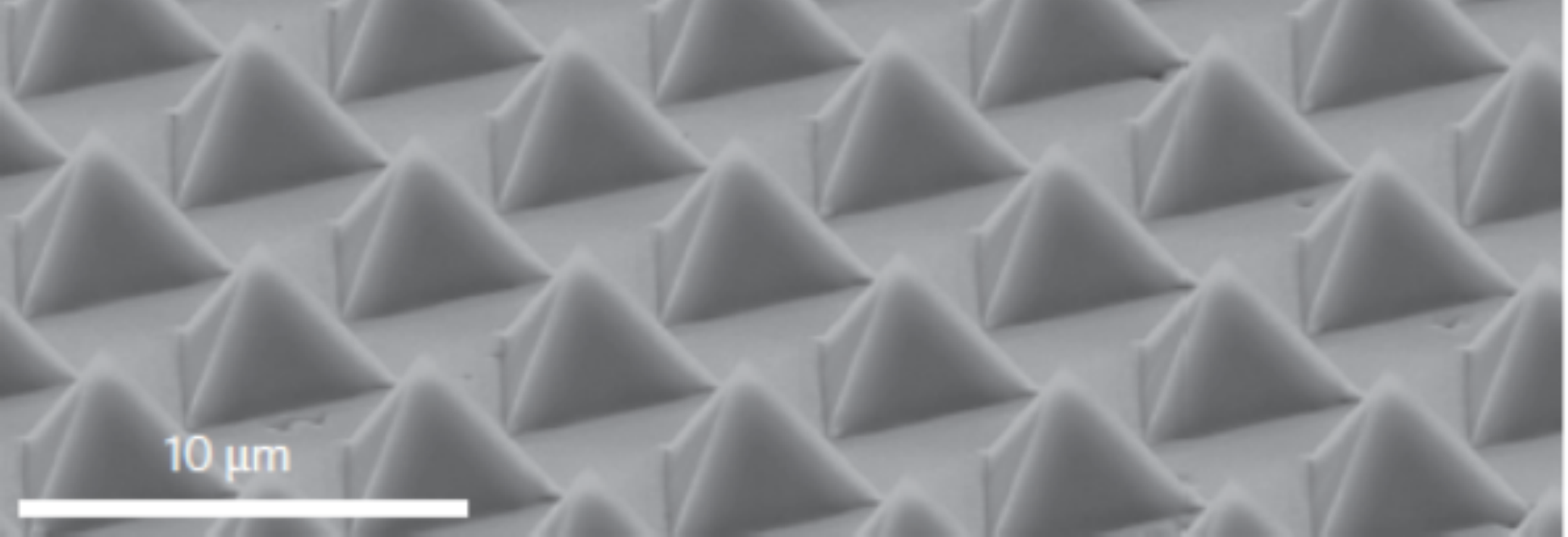}}
\caption{Scanning electron microscopy image of an array of pyramids, grown by MOVPE on periodic recesses in a GaAs surface, and each containing one InGaAsN quantum dot in the apex. (\copyright 2013 NPG, reproduced with permission from~\cite{Juska2013})}
\label{MV4}
\end{figure}

In all these devices, where quantum dots were grown via the Stranski-Krastanow method, the quantum dots were located at random positions. In contrast, Juska \textit{et al.} reported a structure where quantum dots were positioned in a regular array~\cite{Juska2013} (Figure~\ref{MV4}). Their quantum dots are contained in micrometre-size pyramids, which were grown by MOVPE on periodic recesses in a GaAs surface. There are areas of the sample where 15\% of the quantum dots emit entangled photon pairs, with fidelities to a maximally entangled state up to $0.721 \pm 0.043$ without temporal selection. Later, the same group also implemented electrical excitation of entangled-photon emitting pyramidal quantum dots~\cite{Chung2016}. A key element here is the growth of the quantum dots along the $<$111$>$ direction, which leads to a symmetric confining potential of the quantum dots and reduces the fine-structure splitting. In contrast, the regular Stranski-Krastanow growth is prohibited along $<$111$>$, but is performed along $<$100$>$. Growth along $<$111$>$ was also used by Kuroda \textit{et al.}~\cite{Kuroda2013}, who used a droplet epitaxy process to fabricate symmetric quantum dots emitting highly entangled photon pairs, with a fidelity to a maximally entangled state of $0.86 \pm 0.02$ without temporal selection.

\begin{figure}[htb]
\centerline{\includegraphics[width=0.8\textwidth]{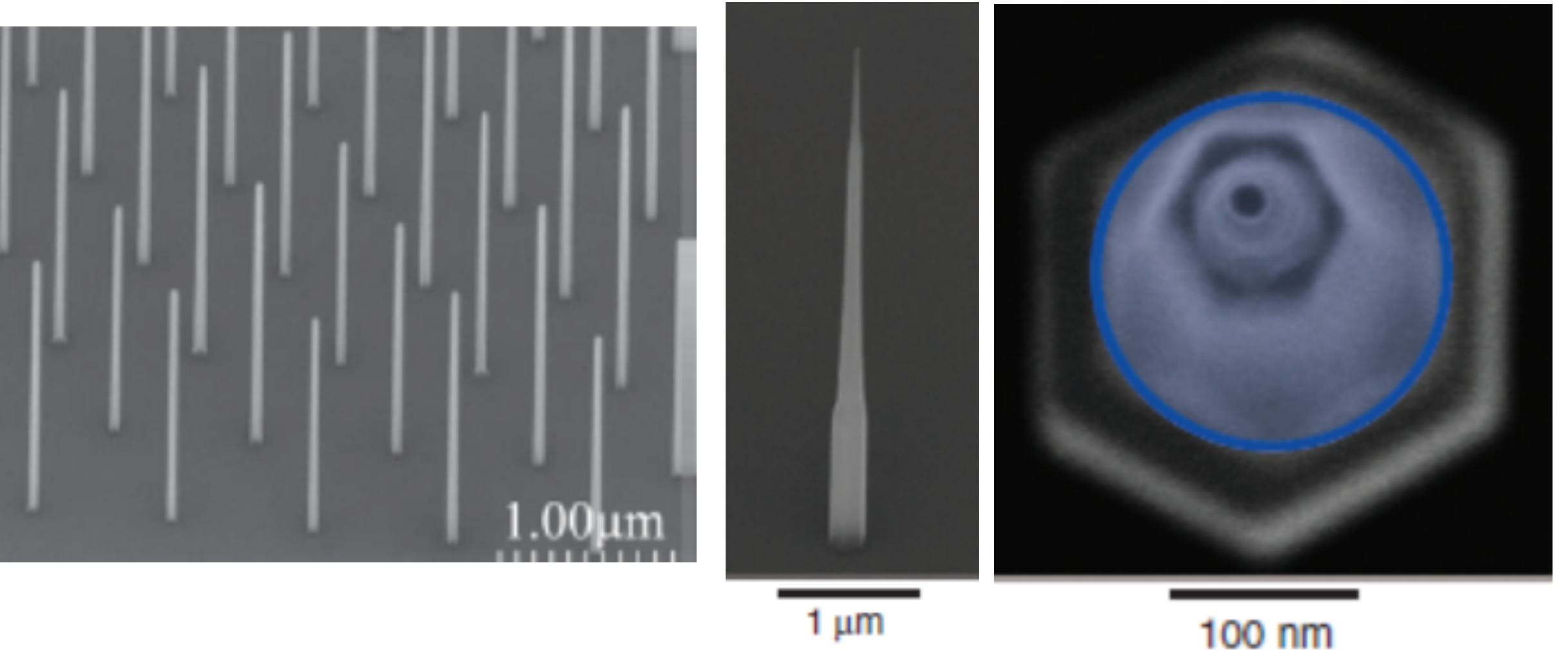}}
\caption{a) Array of InP nanowire waveguides, each containing one InAsP quantum dot. (\copyright 2014 ACS, reproduced with permission from~\cite{Huber2014}) b) The tapered end of one nanowire facilitates the efficient outcoupling of the entangled photons from the quantum dot. (\copyright 2014 under Creative Commons, reproduced from~\cite{Versteegh2014}) c) The wires have an hexagonal shape. (\copyright 2014 under Creative Commons, reproduced from~\cite{Versteegh2014})}
\label{MV5}
\end{figure}

\noindent Position control is also obtained in regular arrays of InP nanowires, containing InAsP quantum dots (Figure~\ref{MV5}). It was theoretically predicted that for $<$111$>$-grown quantum dots in such wires, the fine-structure splitting vanishes~\cite{Singh2009}, and indeed, strong quantum entanglement was observed from nanowire quantum dots~\cite{Versteegh2014,Huber2014}, and it was shown that the emitted photon pairs violate Bell’s inequality~\cite{Jons2015}. An advantage of these nanowires is that the directionality of the emission is ensured by the waveguiding effect of the nanowire, while a tapered end results in the efficient outcoupling of a Gaussian beam. The brightness was measured to be 0.0025 photon pairs per excitation into the first lens, and the maximum fidelity to a maximally entangled state was $0.817 \pm 0.002$ without temporal selection~\cite{Jons2015}. The brightness and entanglement fidelity of various quantum dot and parametric down-conversion sources are compared in~\cite{Jons2015}.

\subsection{Comparison of performance}

A quantitative comparison of all types of entangled photon-pair sources is very challenging. Different scientific communities not only follow different approaches to characterize their sources but also have different applications in mind, thus optimizing their sources in a different way.\\
An important figure of merit for entangled photon-pair sources is obviously their degree of entanglement that can be characterized in many different ways, as shown in Section~3. Currently, parametric down-conversion sources are the ones that produce the higher degree of entanglement, in particular sources based on periodically-poled KTiOPO$_4$ (PPKTP) crystals in Sagnac-loops reach Bell fidelities and purities close to unity~\cite{Pryde2016}. Even though AlGaAs and Si sources based on parametric down-conversion have not yet reached this extremely high level of quality, they have already demonstrated highly entangled states, showing in particular significant violations of Bell's inequalities with measured values of the Bell parameter up to $S \approx 2.7$~\cite{Sarrafi2015,Autebert2016}. Recent results obtained with GaAs quantum dots, grown via the droplet etching method, have shown excellent Bell state fidelities of $0.94 \pm 0.01$ and values of the Bell parameter up to $2.64 \pm 0.01$~\cite{Trotta2016arXiv}. The entanglement measurements reported for these sources are currently limited by technological imperfections, both in the fabrication process of the sources and in the single-photon detectors used to characterize them; on-going technological developments let envision further progress in multiple ways.\\
Another important figure of merit of photon-pair sources is their brightness, one possible definition of which is their probabilty of emitting a photon pair per excitation pulse. Parametric down-conversion sources, because of the thermal statistics of the number of photons they emit, are intrinsically limited to a brightness lower than a few percents at best (0.001 to 0.01 in practice) in order to maintain their high entanglement quality. Conversely, sources based on quantum dots do not present such a limitation and could in principle reach a brightness arbitrary close to unity, and thus generate entangled photon pairs on-demand~\cite{Muller2014}. The brightest entangled photon-pair source reported so far is based on InAs quantum dots embedded in a photonic molecule, reaching a photon-pair efficiency of 0.12 pairs per excitation pulse, with a measured fidelity to a maximally entangled state of 0.63~\cite{Dousse2010}.\\
The emission wavelength is not the same for all processes. Nonlinear semiconductor sources are usually engineered to emit entangled photon pairs in the Telecom C-band around 1550~$nm$ but could emit them at any desired infrared wavelength corresponding to energies below the bandgap energy. Currently quantum dots have shown entanglement at 780~$nm$ (GaAs dots), between 870~$nm$ and 930~$nm$ (InGaAs and InAsP dots) and in the Telecom O-band around 1300~$nm$ (InAs dots).\\
Finally, the operation temperature should also be taken into account. Spontaneous parametric down-conversion sources, including semiconductor ones, work at room temperature. However, quantum-dot-based sources operate at cryogenic temperatures and the highest operation temperatures reported for a quantum dot generating entangled photon pairs so far is 53\,K~\cite{Dousse2010b} with temporal post-selection, and 30\,K without the need for temporal post-selection~\cite{Hafenbrak2007}.

\section{Applications and prospects}

As mentioned in the introduction, practical sources of entangled photon pairs are one of the necessary building blocks of future long-distance quantum communication networks~\cite{Zoller2001,Kimble2008}. Indeed, entangled photon pairs, in association with quantum memories~\cite{Tittel2009,Tittel2013,Heshami2016}, enable the quantum teleportation~\cite{Bennett1993,Bouwmeester1997} of quantum states between arbitrarily distant locations, thus overcoming the problem of propagation losses which limit the reach of direct link transmission. Entangled photon pairs are also needed for device-independent quantum cryptography~\cite{Ekert1991,Acin2006,Pironio2009} which relies on the violation of Bell's inequalities to guarantee the unconditional security of the secret key exchange even when untrusted devices are used. Although this application is still too demanding for present detector technologies, entangled photon-pair sources have already been used in proof-of-principle demonstrations of entanglement-based quantum key distribution. In particular, the broadband emission of an AlGaAs source of polarization-entangled photon pairs has been recently exploited to distribute quantum keys among several pairs of users using standard telecom components~\cite{AutebertPrep}.

Up to now, researchers working on the development of semiconductor-based sources of entangled photons have mainly concentrated their efforts on designing new devices and improving their performances. Thus these sources have not yet been used much for the implementation of quantum communication protocols. However, current efforts also aim at improving their scalability, engineering the emitted quantum states, and further exploiting the integration possibilities offered by the semiconductor platform to fabricate not only sources of quantum photonic states, but also circuits to manipulate them~\cite{Orieux2016}. In the following we give some examples of on-chip manipulation of quantum light that have been demonstrated on the different platforms we discusssed in this review.\\
The linear electro-optical effect of AlGaAs has allowed to demonstrate a tunable Mach-Zehnder interferometer~\cite{Wang2014} that has been used to show two-photon interference with a visibility of 95\% and manipulation of two-photon states with a visibility of 84\%.
On-chip filtering of the pump beam in parametric processes, which is a particularly difficult task in the case of four-wave mixing, has also been addressed: the spectacular advances in CMOS integrated photonics have enabled the integration of filters with Si sources~\cite{Grassani2014}, achieving 95~dB of rejection of the pump light.
As we have seen in this review, different degrees of freedom of the photons can be used to produce entangled states with parametric processes. Depending on the target application, some of these degrees of freedom will be more adapted than others, it is thus important to develop devices able to switch between them. Such converters have been demonstrated on the silicon platform~\cite{Wang2016Bristol,Olislager2013} for transfering entanglement from path to polarization and the other way around. The distribution of high fidelity entanglement has thus been achieved between two integrated silicon quantum chips linked by an optical fibre~\cite{Wang2016Bristol}.\\
Another aspect under development with parametric sources is the implementation of high-dimensional Hilbert spaces (e.g. frequency, optical angular momentum) for quantum communications~\cite{Mair2001}. On the one hand this would increase the channel capacity and provide a more secure key distribution. In addition, implementing a single-photon qudit state would result in a significantly lower power consumption during state preparation, transmission and detection processes.\\
In the case of quantum dot-based sources, efforts are under way to direct the light emission in the transverse plane so as to be able to manipulate the emitted quantum state directly on-chip. An integrated autocorrelator has thus been demonstrated~\cite{Jons2015b}, in which a quantum dot has been embedded in a 50/50 directional coupler.
Another interesting feature of quantum dots is the fact that they possess an optically adressable spin which could be used as a quantum memory. This is why the development of spin-photon interfaces~\cite{Gao2012,DeGreve2012,Schaibley2013} is a key issue towards quantum computing and quantum repeaters. For example, a spin-photon interface based on two orthogonal waveguides, able to map the polarization emitted by a quantum dot to path-encoded photons, has been developed~\cite{Luxmoore2013}. Recently, quantum dots have been used to generate hyper-entanglement, where the photon pairs are entangled in polarization and time-bin~\cite{Prilmuller2017}. In addition, mode-entanglement generated by indistinguishable photons emitted from quantum dots has been used for quantum sensing~\cite{Muller2016,Bennett2016} and boson sampling~\cite{Loredo2016b,He2016,Wang2016b}.

In conclusion, the generation and manipulation of quantum states of light are very exciting topics for both new science and new applications.
The developments in semiconductor optical technologies including quantum light sources, guided-wave circuits, and detectors have opened up promising roads which will yield new modes of communication, sensing, and simulation based on light. The future looks very bright for photonic quantum information technologies!


\section{Acknowledgments}

We thank J.-M. Gérard and V. Zwiller for fruitful discussions.
S.D. acknowledges Agence Nationale de la Recherche (ANR-14-CE26-0029) and Region Ile-de- France in the framework of DIM Nano-K for financial support. M.A.M.V. acknowledges funding from Vetenskapsrådet under grant agreement No. 2016-04527. K.D.J. acknowledges funding from the MARIE SKŁODOWSKA-CURIE Individual Fellowship under REA grant agreement No. 661416 (SiPhoN).
S.D. is member of Institut Universitaire de France.

\section*{References}
\bibliography{RPP_SCsources2017_biblio}
\bibliographystyle{ieeetr}

\end{document}